\documentclass{aa}  

\usepackage{lscape}

\usepackage{graphicx}
\usepackage{txfonts}
\usepackage[colorlinks=true,linkcolor=blue, citecolor=blue, urlcolor=blue, breaklinks=true]{hyperref}
\usepackage{subfig}

\begin{document} 

\title{Precise calibration of the dependence of surface brightness--colour relations on colour and class for late-type stars\thanks{based on CHARA/VEGA observations.}}

\authorrunning{Salsi et al.} 

  \author{A. Salsi\inst{1}, N. Nardetto\inst{1}, D. Mourard\inst{1}, O. Creevey\inst{1},  D. Huber\inst{2}, T. R. White\inst{3, 4, 5}, V. Hocdé\inst{1}, F. Morand\inst{1}, I. Tallon-Bosc\inst{6},  C. D. Farrington\inst{7}, A. Chelli\inst{1}, G. Duvert\inst{8}}

   \institute{Université Côte d’Azur, OCA, CNRS, Laboratoire Lagrange, France\\
              \email{anthony.salsi@oca.eu}
         \and
                  Institute for Astronomy, University of Hawaii, 2680 Woodlawn Drive, Honolulu, HI 96822, USA
         \and
         Sydney Institute for Astronomy, School of Physics A28, The University of Sydney, NSW 2006, Australia
         \and
         Stellar Astrophysics Centre, Department of Physics and Astronomy, Aarhus University, DK-8000 Aarhus C, Denmark
         \and
         Research School of Astronomy and Astrophysics, Mount Stromlo Observatory, The Australian National University, Canberra, ACT 2611, Australia
         \and
                  Univ. Lyon, Univ. Lyon 1, Ens de Lyon, CNRS, Centre de Recherche Astrophysique de Lyon UMR5574, F-69230 Saint-Genis-Laval, France
        \and
         The CHARA Array, Mount Wilson Observatory, Mount Wilson, CA91023 USA
         \and
         Univ. Grenoble Alpes, CNRS, IPAG, 38000 Grenoble, France
         \\
             }

   \date{Received... ; accepted...}

% \abstract{}{}{}{}{} 
% 5 {} token are mandatory
 
  \abstract
  % context heading (optional)
  % {} leave it empty if necessary  
   {Surface brightness--colour relations (SBCRs) are used to derive the stellar angular diameters from photometric observations. They have  various astrophysical applications, such as the distance determination of eclipsing binaries or the determination of exoplanet parameters. However, strong discrepancies between the SBCRs still exist in the literature, in particular for early and late-type stars.}
   {We aim to calibrate new SBCRs as a function of the spectral type and the luminosity class of the stars. Our goal is also to apply homogeneous criteria to the selection of the reference stars and in view of compiling an exhaustive and up-to-date list of interferometric late-type targets.}
   {We implemented criteria to select measurements in the JMMC Measured Diameters Catalog (JMDC). We then applied additional criteria on the photometric measurements used to build the SBCRs, together with stellar characteristics diagnostics.}
   {We built SBCRs for F5/K7-II/III, F5/K7-IV/V, M-II/III and M-V stars,
   with respective RMS of $\sigma_{F_{V}} = 0.0022$ mag, $\sigma_{F_{V}} = 0.0044$ mag, $\sigma_{F_{V}} = 0.0046$ mag, and $\sigma_{F_{V}} = 0.0038$ mag. This results in a precision on the angular diameter of 1.0\%, 2.0\%, 2.1\%, and 1.7\%, respectively. These relations cover a large $V-K$ colour range of magnitude, from 1 to 7.5. Our work demonstrates that SBCRs are significantly dependent on the spectral type and the luminosity class of the star. Through a new set of interferometric measurements, we demonstrate the critical importance of the selection criteria proposed for the calibration of SBCR. Finally, using the Gaia photometry for our samples, we obtained (G-K) SBCRs with a precision on the angular diameter between 1.1\% and 2.4\%.}
   {By adopting a refined and homogeneous methodology, we show that the spectral type and the class of the star should be considered when applying an SBCR. This is particularly important in the context of PLATO.}

   \keywords{stars: fundamental parameters -- cosmology: distance scale -- techniques: interferometric 
               }

   \maketitle
   
%
%-------------------------------------------------------------------

\section{Introduction}\label{Intro}

 \begin{figure*}
   \centering
   \includegraphics[width=0.9\hsize]{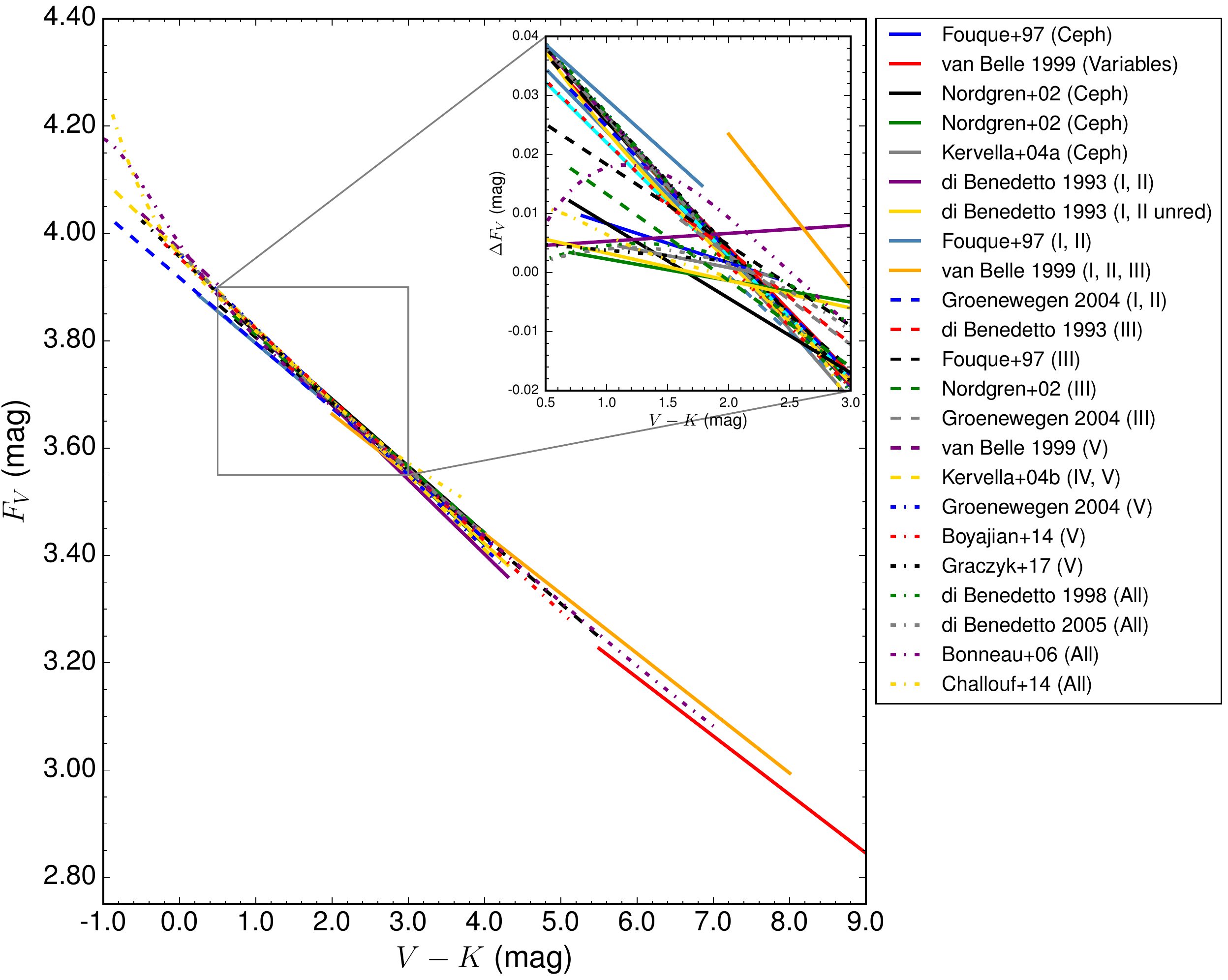}
      \caption{The 23 SBCRs in the literature are plotted as a function of the $V-K$ colour (over their validity domain), and comparatively to the \cite{2004AA...426..297K} relation in $\Delta F_V$ on the top right corner, between 0.5 and 3.0 mag, which is taken as a reference (for clarity). We note, however, that the \citet{2004AA...426..297K} relation is in principle valid only over the -0.85 to 4.10 $V-K$ range. The references for these SCBRs are as follows: \cite{Fouque1997,vB99}; \citet{Nordgren02}; \citet{Kervella04a}; \citet{dB93}; \citet{G04}; \citet{2004AA...426..297K}; \citet{Boyajian14};  \citet{Graczyk17}; \citet{dB98}; \citet{dB05}; \citet{Bonneau06}; \citet{Challouf}.}
         \label{all_SBCRs}
 \end{figure*}

Surface brightness--colour relations (SBCRs) are very convenient tools for easily estimating the angular diameter of a star from photometric measurements. For instance, the SBCR plays a central role in the distance determination of eclipsing binaries, by combining the linear diameter (derived from light curve and velocimetry) and the estimated angular diameter of their components.

Recently, in the course of the Araucaria project \citep{Araucaria}, \citet{Pietr} used this method to constrain the Large Magellanic Cloud distance to $1\%$. The PLATO (PLAnetary Transits and Oscillations of stars, \citet{PLATO}) space mission, planned for launch in 2026, will characterise exoplanetary systems, with the transit method. PLATO will thus provide the ratio of stellar-to-planet radii with 1\% precision, while an SBCR combined with Gaia parallaxes will give access to the stellar radius. 

So far, 23 SBCRs have been established, covering all spectral types and luminosity classes. \citet{HDR_Nico} compares these SBCRs, and shows that they are precise but inconsistent for late-type stars (at the 10\% level), while they are rather imprecise for early-type stars (around 7\% precision, \citet{Challouf}). Besides this, several studies, such as \citet{Fouque1997} and \citet{2004AA...426..297K} point out a significant difference in the SBCRs according to the luminosity class of the stars (see also, \citet{HDR_Nico}). They also suggest the impact of the activity of the star. \citet{Chelli2016} also proposed a different method based on so-called pseudo-magnitudes to build the JSDC catalogue, including 450K star diameters.

In the present work, we restrict our analysis to late-type stars, following the PLATO specifications, taking into account the luminosity classes as suggested by previous studies mentioned above. We only consider stars from F5 to K7, which corresponds to an effective temperature ($T_{\mathrm{eff}}$) lower  than $6510K$ \citep{Pecaut} (or $V-K \geq 1$ mag) and higher than $4050K$. We also consider $\log g = 4.0$ as the typical separation between dwarfs (V), sub-giants (IV) on one side, and giants (III) on the other side. This leads to four working samples; F5/K7 giants, F5/K7 sub-giants and dwarfs (II/III and IV/V luminosity classes, respectively), M giants, and M sub-giants and dwarfs.

We first present the SBCRs existing in the literature in Sect. \ref{def_SBCRs}. We then describe the selection of our interferometric and photometric measurements in Sect. \ref{Selection}, as well as the reddening law we used to correct the interstellar extinction. Our calibrated SBCRs are presented in Sect. \ref{new_SBCRs} and discussed in Sect. \ref{discussion}. 

\begin{table*}
\caption{Description of the stellar characteristics (top part), interferometric (middle part), and photometric (bottom) criteria we considered for the data selection. Right column shows labels relative to these criteria, which we included as a note in the final samples of stars (see Table \ref{sample_table}).}             % title of Table
\label{criteria_table}   
\centering                         
\begin{tabular}{c c c}       
\hline\hline        
& Criterion & Label \\   
\hline                      
%\begin{flushleft}
Stellar characteristics (see Sect. \ref{criteria_section})
%\end{flushleft} 
& Variable\tablefootmark{a} & V\\       
& Spectroscopic binary & SB\\
& Multiple & M\\
& Doubt on luminosity class & LumC \\
& Semi-Regular pulsating star & Sr Puls\\       
& Fast Rotator & FRot \\
\hline
%\begin{flushleft}
Interferometric (see Sect. \ref{interfero_criteria}) 
%\end{flushleft} 
& Not compatible data and no visibility curve & NVisC\\
& 8-13 um band & 8-13\\
& Data very far from the general trend & Bad\\
& High visibility measurements ($V^2$ > 0.8) & hVis\\
& Excellent visibility curve in the other reference\tablefootmark{b} & eVisC\\
& Large visible band problem & LvisBand\\
\hline
%\begin{flushleft}
Photometric (see Sect. \ref{inf_photo}) 
%\end{flushleft} 
& High $K$ magnitude uncertainty & hK\\
\hline                                  
\end{tabular}
\tablefoot{
\tablefoottext{a}{BY: BY Dra type, TT: T-Tauri type, RS: RS CVn type, dS: delta Scuti type, Cep: Cepheids.}
\tablefoottext{b}{In case of inconsistent redundancies.}
}\label{table_criteria}
\end{table*}

\section{Definition and surface brightness--colour relations in the literature}\label{def_SBCRs}

The surface brightness of a star is the flux density emitted per unit angular area. The Stefan-Boltzmann law connects the surface brightness to the effective temperature $T_{\mathrm{eff}}$. An empirical relation between the effective temperature and the colour (i.e. the difference in magnitude measured in two different spectral bands) of the star is then found to relate the surface brightness to the colour. The first historical definition of the surface brightness was established by \citet{SBCR_Wesselink}, depending on the bolometric correction and the effective temperature of the star. \citet{SBCR_Wesselink} then used this definition to show the correlation between the surface brightness and the colour of the star. Later, \citet{Barnes1976} built another definition of the surface brightness, noted $F_{\lambda}$, written as follows:

\begin{equation}\label{4.2207}
F_{\lambda} = C - 0.1 m_{\lambda_{0}} - 0.5 \log {\theta_{\mathrm{LD}}},
\end{equation}

where $\theta_{\mathrm{LD}}$ is the limb-darkened angular diameter of the star, $m_{\lambda_{0}}$ is the apparent magnitude corrected from the interstellar extinction, and $C$ is a constant. After \citet{Fouque1997}, $C$ depends on the Sun bolometric magnitude $M_{\mathrm{bol}_{\odot}}$, its total flux $f_{\odot,}$ and the Stefan-Boltzmann constant $\sigma$ through the following relationship \citep{Fouque1997}:

\begin{equation}\label{coeff_C}
C = 0.1 M_{\mathrm{bol}_{\odot}} + 1 + 0.25 \log \frac{4 f_{\odot}}{\sigma},
\end{equation}

and it is found to be equal to $4.2207$. More recent and accurate estimations of solar parameters \citep{Mamajek, Prsa} lead to a slightly different value, 4.2196, which we took for our study. The definition of the surface brightness can be rewritten as follows:

\begin{equation}\label{new_def_SBCR}
F_{\lambda} = 4.2196 - 0.1 m_{\lambda_{0}} - 0.5 \log {\theta_{\mathrm{LD}}}.
\end{equation}

On the other hand, the bolometric surface flux $f_{\mathrm{bol}}$ of a star, which is expressed as the ratio between the bolometric flux $F_{\mathrm{bol}}$ and the squared limb-darkened angular diameter $\theta_{\mathrm{LD}}^2$, is linearly proportional to its effective temperature $T_{\mathrm{eff}}^4$. It is thus also linearly linked to the colour $m_{\lambda_{1}}-m_{\lambda_{2}}$. In this way, the surface brightness can be estimated by the following linear relation:

\begin{equation}\label{SBCR_def}
F_{\lambda_{1}} = a \left(m_{\lambda_{1}}-m_{\lambda_{2}}\right) + b.
\end{equation}

The previous equation corresponds to the so-called surface brightness--colour relation (SBCR). By injecting Eq. \ref{new_def_SBCR} into Eq. \ref{SBCR_def}, the SBCR allows us to directly estimate the limb-darkened angular diameter of the star. We used this definition of the SBCR in this work. \citet{HDR_Nico} demonstrated the existence of various definitions of the SBCR in the literature. By carrying out suitable conversions, we compare the 23 SBCRs in Fig. \ref{all_SBCRs}, as a function of the $V-K$ colour. In the following, we consider the ($V$, $V-K$) colour system, as it is known to provide the lowest dispersion in the SBCRs \citep{2004AA...426..297K}. As shown by the figure, the SBCRs in the literature are rather consistent around $V-K$ = 2 mag, with an expected precision on the derived angular diameter (using any SBCR) of about 2\%. However, some discrepancies are clear on the outer edges of the surface brightness versus $V-K$ colour diagram, as already mentioned. In order to calibrate the SBCRs, we need the $V$ and $K$ magnitudes, the limb-darkened angular diameter, an extinction law, as well as diagnostics on star activity. We describe the strategy we implemented to find such information in the next sub-sections.

\section{Methodology and selection criteria}\label{Selection}

The quality and robustness of an SBCR is strongly related to the definition of the samples of stars used for its calibration and to the correct explanation of its domain of validity. In this section, we present the method employed to define our samples on the basis of the JMDC catalogue, and we detail the various selection criteria that were developed. 

\subsection{JMDC catalogue}\label{JMDC}

The most complete and up-to-date catalogue that lists all the interferometric measurements that have been done so far is the JMMC Measured stellar Diameters Catalog\footnote{Available on the VizieR database at  \url{https://vizier.u-strasbg.fr/viz-bin/VizieR?-source=II/345}.} \citep{Duvert}. As of February 2020, this catalogue contains 1672 rows. Among all these measurements, the current number of individual stars with observed diameters is 885. The catalogue lists the uniform disc angular diameter $\theta_{\mathrm{UD}}$, the limb-darkened angular diameter $\theta_{\mathrm{LD,}}$ and the $\theta_{\mathrm{UD}}$ to $\theta_{\mathrm{LD}}$ conversion factor $\mu_{\lambda}$ if available. A "notes" column is included and contains some information about the star. The observing technique is indicated: optical interferometry, lunar occultation or intensity interferometry. We cross-matched the \textit{Simbad} database with the JMDC catalogue to obtain photometric information (see Sect. \ref{visible_part}).

\begin{figure*}
   \centering
   \includegraphics[width=\hsize]{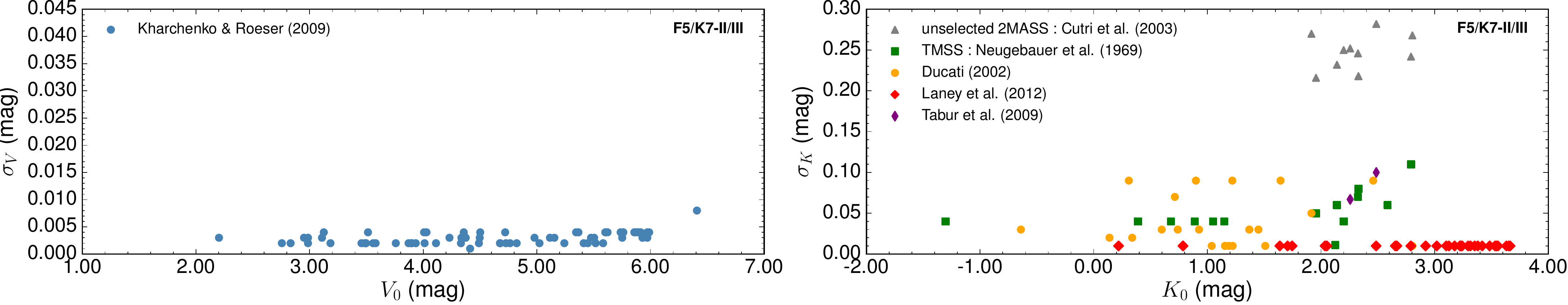}
   \includegraphics[width=\hsize]{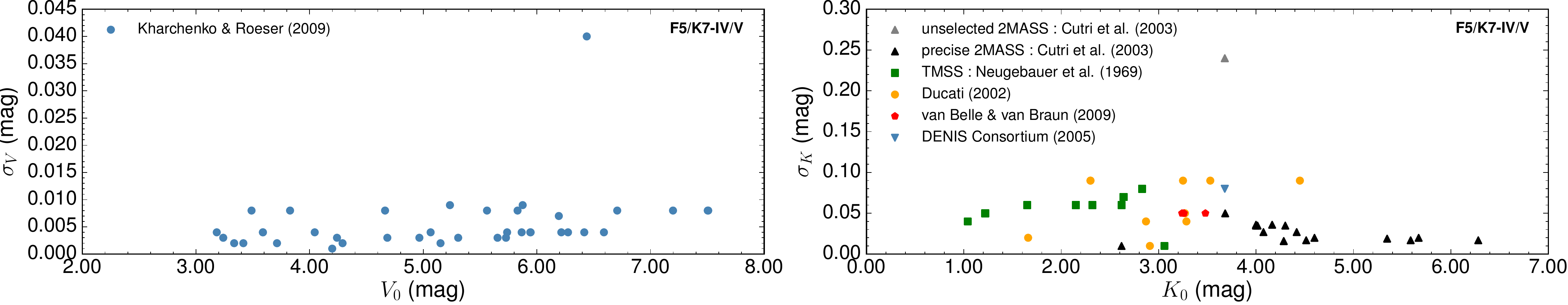}

   \includegraphics[width=\hsize]{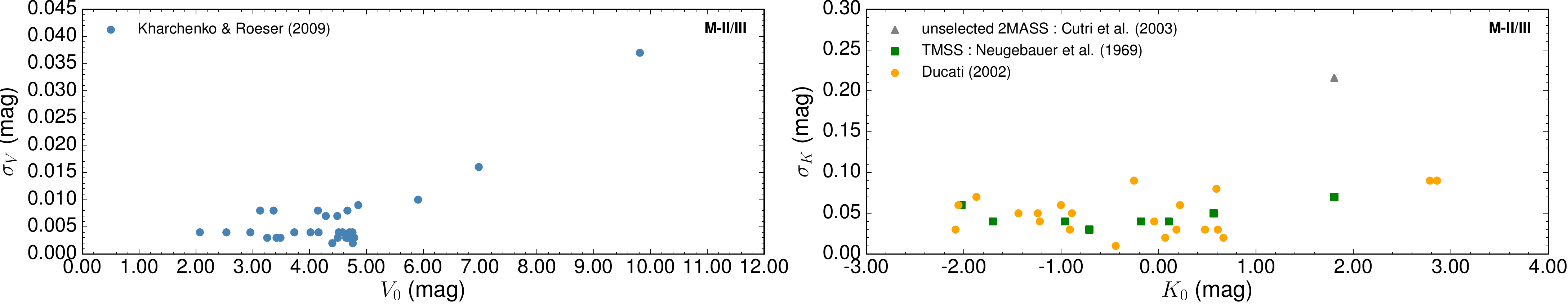}
   \includegraphics[width=\hsize]{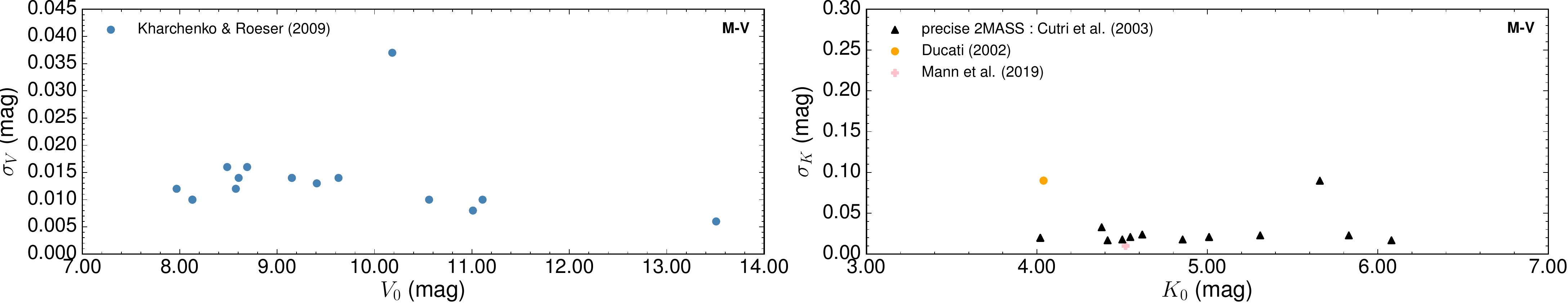}

      \caption{$\sigma_V$ vs. $V_0$ (left panels) and $\sigma_K$ vs. $K_0$ (right panels) plotted for the four samples, indicated in the top-right corner of each graphic. The photometric sources are noted in the legend.}
         \label{mags_ref}
 \end{figure*}
 
\subsection{Common criteria}\label{common_criteria}

To build SBCRs, one needs several input data; $\theta_{\mathrm{LD}}$, $\sigma_{\theta_{\mathrm{LD}}}$, $V$, $\sigma_V$, $K,$ and $\sigma_K$. We list the general criteria applied to our samples: (i) consider the spectral type (later than F5) and the luminosity class (II, III, IV or V) of the star; (ii) retain only optical interferometry measurements; (iii) reject measurements without all the necessary data ($\theta_{\mathrm{LD}}$, $\sigma_{\theta_{\mathrm{LD}}}$, $V$, $\sigma_V$, $K$ and $\sigma_K$).

\subsection{Stellar characteristics criteria}\label{criteria_section}

We implemented six more criteria based on the characteristics of the star. These criteria are presented in the top part of Table \ref{table_criteria}, with their corresponding labels used in the final table. When a star has one of these activity signs, it is not used to constrain the SBCRs, but it still appears in our final table of parameters. However, we needed to make several exceptions in the selection process. Among the remaining stars, the variability was prevalent in the F5/K7 giants sample. This criterion is thus not considered when selecting giants. We quantitatively study this point later in Sect. \ref{discussion_1}. Moreover, given the very low number of M dwarf measurements, no selection is based on their activity (only the quality of the interferometric data, see below).

\subsection{Interferometric criteria}\label{interfero_criteria}

To build accurate SBCRs, one needs precise angular diameter measurements. We arbitrarily excluded measurements with errors on the angular diameter larger than $8\%$. We then removed measurements done in the $8-13$ $\mu m$ band to avoid the contamination of the flux of the star by any materials, like a circumstellar envelope or dust. In some cases, we find data that is totally inconsistent (more than 5$\sigma$) with the SBCRs, due to inaccurate conversions from $\theta_{\mathrm{UD}}$ to $\theta_{\mathrm{LD}}$, bad observation quality and/or poor spatial frequency coverage in the visibility curve. The corresponding data are then flagged as "NVisC", "Bad" or "LvisBand" in Table \ref{table_criteria}. If a star has several interferometric independent measurements (e.g. on different instruments) satisfying all the criteria, we keep them all in the sample.

The LD diameters in the JMDC are predominantly deduced from the measured UD diameters using Claret's grids \citep{Claret1995, Claret2000, Claret}. Claret's grids have a step of 250 K in temperature, thus the largest error we can make on the temperature is 125 K without any interpolation. As mentioned by \citet{Nardetto2020} in a recent work, this error on the temperature leads to an error of 0.3\% on all angular diameters, well below the typical errors of our samples. Moreover, the angular diameters computed with SATLAS \citep{SATLAS2008, SATLAS2013} are 0.4\% larger from those deduced with \citet{Claret} grids for K giants. For dwarfs, we expect an even smaller difference. This means that even if the UD to LD conversion is not done homogeneously on our JMDC samples, the impact on the value of the angular diameter is well below the quoted uncertainty.

\subsection{Visible photometry}\label{visible_part}

Surface brightness--colour relations are strongly dependent on the photometry used for the calibration. We thus took into consideration both $V$ and $K$ uncertainties to properly build our SBCR fitting strategy. We considered visible magnitudes from the \citet{Kharchenko} catalogue. This catalogue gathers measurements from several other catalogues (Hipparcos-Tycho catalogues, Carlsberg Meridian Catalog and the Positions and Proper Motions catalogue), and all the visible magnitudes are given in the Johnson V filter. The strong interest of this catalogue is the accuracy of the measurements, with an error on the visible magnitude rarely exceeding 0.01 mag (see Fig. \ref{mags_ref}).

\subsection{Infrared photometry and additional criterion}\label{inf_photo}

The uniformity of the infrared $K$ magnitude was more complicated to fulfill since the 2MASS catalogue \citep{2MASS}, which is the most complete catalogue of infrared photometry, is not very accurate for a lot of the measurements (mainly because of saturation issues). We decided to consider only infrared measurements with an error below 0.15 mag. For stars with a precision larger than 0.15 magnitude on the 2MASS photometry, we searched other catalogues for more accurate infrared measurements. This allowed us to keep 10 additional stars, indicated by grey triangles in Fig.~\ref{mags_ref}, for which we found more precise infrared photometry. The various sources we found for the infrared photometry are given in the legend in Fig. \ref{mags_ref}. They are also listed in Table \ref{table_photometry} with their corresponding labels. This induced a new selection criterion, labelled as "hK". 

Among all the catalogues we use for the infrared K-photometry, only \citet{Ducati} and \citet{TMSS} use Johnson photometry without conversion into 2MASS photometry. This corresponds to 85 stars over the 153 in our samples. 
%After a deep research in the literature, we did not find any reference that proposes a Johnson-K to $Ks$ conversion. 
We did a test by considering only $Ks$ photometries to constrain our SBCRs. We find a consistency of less than 1-$\sigma$ between the one with only 2MASS photometry and the other one with heterogeneous photometry. To evaluate the impact of the heterogeneous infrared photometry, we compared both photometries for 4 stars in our samples: HD140283, HD3651, HD4628, and HD75732. We found a difference of 0.05\%, 0.35\%, 2.5\%, and 1.2\%, respectively, leading to a difference of 0.1\%, 0.7\%, 4.5\%, and 2.8\% on the angular diameter. Both $K$ and $K_s$ photometries are consistent in the error bars for these four stars. The difference is therefore minimal, provided that $K$ and $K_s$ photometries differ within 2\%. To conclude, our SBCRs are mixed with 2MASS/Johnson $-K$ photometries, but both are consistent, meaning that our SBCRs can be used with the two photometries without including any significant bias on the angular diameter.

\begin{table}
\caption{Infrared photometry sources with their corresponding labels included in Table \ref{table_criteria}.}           
\centering                          % used for centering table
\begin{tabular}{c c}        % centered columns (4 columns)
\hline\hline                 % inserts double horizontal lines
Infrared photometry source & Label \\    % table heading 
\hline                        % inserts single horizontal line
TMSS \citep{TMSS} & T \\
\citet{Ducati} & Du \\
2MASS \citep{2MASS} & 2M \\
\citet{Denis} & De \\
\citet{vanBelle} & V \\
\citet{Tabur} & Ta \\
\citet{Laney2012} & La\\ 
\citet{Mann2019} & M19 \\
\hline                                  
\end{tabular}\label{table_photometry}
\end{table}

%\longtab{
\begin{sidewaystable*}
%\begin{longtable}{cccccccccccccc}
\caption{Part of the F5/K7 giants sample after applying common criteria. The "Notes" column contains keywords relative to stellar characteristics, interferometric and photometric criteria when the star is not considered. Refer to Table \ref{table_criteria} for a description of these keywords. "K ref" column includes infrared photometry sources as listed in Table \ref{table_photometry}.}
\centering
\begin{tabular}{cccccccccccccc}
\hline
\hline
Name    &       $\theta_{\mathrm{LD}}$  &       $\sigma_{\theta_{\mathrm{LD}}}$ &       Band    &       Source  &       Sp. Type    &       $V$     &       $\sigma_V$      &       $K$     &       $\sigma_K$      &       $A_{\lambda_{V}}$       &       $K$ ref.    &       Notes   \\
        &       [mas]   &       [mas]   &               &               &               &       [mag]   &       [mag]   &       [mag]   &       [mag]   &       [mag]   &               &               \\
\hline
HD100407        &        2.394  &        0.029  &        K      &       \citet{2005AA...436..253T}      &        G7IIIb  &       3.535   &       0.002   &       1.47    &       0.252   &       0.0031  &       2M      &       M       \\
HD10142         &        0.964  &        0.0060         &        H      &       \citet{2018AA...616A..68G}      &        K0III   &       5.933   &       0.003   &       3.557   &       0.010   &       0.0031  &       La      &               \\
HD102328        &        1.606  &        0.0060         &        K'     &       \citet{2010ApJ...710.1365B}     &        K2.5IIIbCN1     &       5.254   &       0.002   &       2.488   &       0.100   &       0.0124  &       Ta      &               \\
HD103605        &        1.098  &        0.0090         &        K'     &       \citet{2010ApJ...710.1365B}     &        K1III   &       5.827   &       0.002   &       3.102   &       0.298   &       0.0248  &       2M      &       hK      \\
HD10380         &        2.81   &        0.03   &        740nm  &       \citet{1999AJ....118.3032N}     &        K3III   &       4.435   &       0.003   &       1.356   &       0.307   &       0.031   &       2M      &       NVisC   \\
HD104985        &        1.032  &        0.022  &        2.15   &       \citet{2008ApJ...680..728B}     &        G8.5IIIb        &       5.785   &       0.003   &       3.273   &       0.303   &       0.0217  &       2M      &       M       \\
HD106574        &        1.498  &        0.027  &        K'     &       \citet{2010ApJ...710.1365B}     &        K2III   &       5.721   &       0.003   &       2.94    &       0.08    &       0.0403  &       T       &       hVis    \\
HD108381        &        2.179  &        0.057  &        550..850nm     &       \citet{2018AJ....155...30B}     &        K1IIIFe0.5      &       4.339   &       0.003   &       1.809   &       0.250   &       0.0031  &       2M      &       Bad + LvisBand      \\
HD11092         &        2.797  &        0.019  &        H or K         &       \citet{2009MNRAS.394.1925V}     &        K4Ib-IIa        &       6.548   &       0.004   &       1.784   &       0.239   &       1.2679  &       2M      &       LumC    \\
HD113049        &        0.971  &        0.021  &        K'     &       \citet{2010ApJ...710.1365B}     &        K0III   &       5.993   &       0.004   &       3.658   &       0.312   &       0.0403  &       2M      &       hK      \\
HD113226        &        3.17   &        0.02   &        740nm  &       \citet{1999AJ....118.3032N}     &        G8III-IIIb      &       2.836   &       0.002   &       0.786   &       0.010   &       0.0031  &       La      &       eVisC   \\
HD113226        &        3.283  &        0.033  &        800nm  &       \citet{2003AJ....126.2502M}     &        G8III-IIIb      &       2.836   &       0.002   &       0.786   &       0.010   &       0.0031  &       La      &               \\
HD113226        &        3.318  &        0.013  &        550..850nm     &       \citet{2018AJ....155...30B}     &        G8III-IIIb      &       2.836   &       0.002   &       0.786   &       0.010   &       0.0031  &       La      &       Bad + LvisBand      \\
HD113996        &        3.09   &        0.019  &        550..850nm     &       \citet{2018AJ....155...30B}     &        K5-III  &       4.782   &       0.002   &       1.2     &       0.09    &       0.0155  &       Du      &       Bad + LvisBand      \\
HD118904        &       1.871   &       0.032   &       K'      &       \citet{2010ApJ...710.1365B}     &       K2III   &       5.493   &       0.003   &       2.69    &       0.07    &       0       &       T       &       Bad     \\
HD11977 &       1.528   &       0.013   &       H       &       \citet{2018AA...616A..68G}      &       G5III   &       4.684   &       0.002   &       2.486   &       0.01    &       0.0062  &       2M      &               \\
HD120477        &        4.691  &        0.022  &        550..850nm     &       \citet{2018AJ....155...30B}     &        K5.5III         &       4.041   &       0.004   &       0.34    &       0.02    &       0.0124  &       Du      &       Bad + LvisBand      \\
HD120477        &        4.72   &        0.04   &        740nm  &       \citet{1999AJ....118.3032N}     &        K5.5III         &       4.041   &       0.004   &       0.34    &       0.02    &       0.0124  &       Du      &               \\
HD12438         &        1.091  &        0.016  &        H      &       \citet{2018AA...616A..68G}      &        G8III   &       5.348   &       0.004   &       3.176   &       0.010   &       0.0031  &       La      &               \\
HD124897        &        20.95  &        0.2    &        2.2    &       \citet{1986AA...166..204D}      &        K1.5IIIFe-0.5   &       0.085   &       0.009   &       -2.911  &       0.170   &       0       &       2M      &       M       \\
HD124897        &        21.373         &        0.247  &        800nm  &       \citet{2003AJ....126.2502M}     &        K1.5IIIFe-0.5   &       0.085   &       0.009   &       -2.911  &       0.170   &       0       &       2M      &       M       \\
HD124897        &        21.6   &        1.2    &        2.2    &       \citet{1983ApJ...268..309D}     &        K1.5IIIFe-0.5   &       0.085   &       0.009   &       -2.911  &       0.170   &       0       &       2M      &       M       \\
HD127665        &        3.901  &        0.0080         &        550..850nm         &       \citet{2018AJ....155...30B}     &        K3-III         &       3.571   &       0.002   &       0.6     &       0.03    &       0.0124  &       Du      &               \\
HD12929         &        6.792  &        0.043  &        700 nm         &       \citet{2016ApJS..227....4H}     &        K2-IIIbCa-1     &       2.01    &       0.004   &       -0.782  &       0.175   &       0       &       2M      &       M       \\
HD12929         &        6.827  &        0.068  &        800nm  &       \citet{2003AJ....126.2502M}     &        K2-IIIbCa-1     &       2.01    &       0.004   &       -0.782  &       0.175   &       0       &       2M      &       M       \\
HD12929         &        6.85   &        0.9    &        800nm  &       \citet{1991AJ....101.2207M}     &        K2-IIIbCa-1     &       2.01    &       0.004   &       -0.782  &       0.175   &       0       &       2M      &       M       \\
HD12929         &        6.88   &        0.03   &        740nm  &       \citet{1999AJ....118.3032N}     &        K2-IIIbCa-1     &       2.01    &       0.004   &       -0.782  &       0.175   &       0       &       2M      &       M       \\
HD12929         &        7.6    &        0.9    &        550nm  &       \citet{1983AA...120..263F}      &        K2-IIIbCa-1     &       2.01    &       0.004   &       -0.782  &       0.175   &       0       &       2M      &       M       \\
HD12929         &        8.7    &        0.5    &        674nm  &       \citet{1989ApJ...340.1103H}     &        K2-IIIbCa-1     &       2.01    &       0.004   &       -0.782  &       0.175   &       0       &       2M      &       M       \\
HD131873        &        10.301         &        0.103  &        800nm  &       \citet{2003AJ....126.2502M}     &        K4-III  &       2.063   &       0.003   &       -1.286  &       0.203   &       0.0031  &       2M      &       M       \\
HD131873        &        8.9    &        1.0    &        550nm  &       \citet{1983AA...120..263F}      &        K4-III  &       2.063   &       0.003   &       -1.286  &       0.203   &       0.0031  &       2M      &       M       \\
HD133124        &        3.055  &        0.077  &        550..850nm     &       \citet{2018AJ....155...30B}     &        K4III   &       4.79    &       0.002   &       1.23    &       0.010   &       0.062   &       Du      &               \\
HD133208        &        2.477  &        0.065  &        800nm  &       \citet{2003AJ....126.2502M}     &        G8IIIaFe-0.5    &       3.476   &       0.002   &       1.223   &       0.165   &       0.0248  &       2M      &       M       \\
HD133208        &        2.484  &        0.0080         &        550..850nm         &       \citet{2018AJ....155...30B}     &        G8IIIaFe-0.5   &       3.476   &       0.002   &       1.223   &       0.165   &       0.0248  &       2M      &       Bad + LvisBand      \\
HD13468         &        0.886  &        0.01   &        H      &       \citet{2018AA...616A..68G}      &        K0III   &       5.932   &       0.004   &       3.666   &       0.010   &       0.0186  &       La      &               \\
HD135722        &        2.74   &        0.02   &        740nm  &       \citet{1999AJ....118.3032N}     &        G8IIIFe-1       &       3.466   &       0.002   &       1.19    &       0.01    &       0.0093  &       Du      &       eVisC   \\
HD135722        &        2.764  &        0.03   &        800nm  &       \citet{2003AJ....126.2502M}     &        G8IIIFe-1       &       3.466   &       0.002   &       1.19    &       0.01    &       0.0093  &       Du      &               \\
HD1367  &        0.754  &        0.013  &        H      &       \citet{2016AA...586A..94L}      &        K0III   &       6.177   &       0.004   &       4.224   &       0.263   &       0.0155  &       2M      &       hK      \\
HD136726        &       2.149   &       0.023   &       550..850nm      &       \citet{2018AJ....155...30B}     &       K4III   &       5.013   &       0.003   &       1.92    &       0.05    &       0.0341  &       T       &               \\
HD136726        &       2.336   &       0.020   &       K'      &       \citet{2010ApJ...710.1365B}     &       K4III   &       5.013   &       0.003   &       1.92    &       0.05    &       0.0341  &       T       &       eVisC   \\
HD13686         &        1.581  &        0.063  &        H or K         &       \citet{2009MNRAS.394.1925V}     &        K2.5Ib-II       &       7.03    &       0.004   &       2.709   &       0.335   &       1.891   &       2M      &       LumC    \\
... & ... & ... & ... & ... & ... & ... & ... & ... & ... & ... & ... & ... \\
\hline
\end{tabular}\label{sample_table}
\end{sidewaystable*}

\subsection{Reddening corrections}

We used the \textit{Stilism}\footnote{The online tool is available at \url{http://stilism.obspm.fr}} online tool \citep{Stilism1, Stilism2} to compute the colour excess $E (B-V)$. This tool produces tridimensional maps of the local interstellar matter (ISM) based on measurements of starlight absorption by dust (reddening effects) or gaseous species. By definition, the interstellar attenuation $A_{V}$ in the visible band is given by

\begin{equation}
A_{V} = R_V \times E(B-V),
\end{equation}

where $R_V$ is the ratio of total to selective absorption in the visible band, for which we adopted $R_V = 3.1$, which corresponds to the typical value in the diffuse ISM \citep{Cardelli}. We then used $A_{K} = 0.119 \times A_{V}$, according to \citet{reddening}.

It is well known that the SBCR is not significantly sensitive to the reddening correction, since the magnitude absorption is compensated by the colour extinction. The visual absorption of our samples rarely exceeds 0.1 mag. To quantify its contribution, we increased the value of the visual extinction on a few stars of our F5/K7 giants sample. A high value $A_V = 0.1$ mag yields to a difference of 0.3\% on the surface brightness, and 0.35\% on the resulting angular diameter. \citet{Nardetto2020} did a test by varying the visible absorption $A_V$ on their entire sample. They find that for a larger absorption of 0.1 mag, the zero-point of their SBCR increases by 0.045 mag (i.e. 0.0045 mag in the $F_V$ definition), which roughly corresponds to the RMS of their relation.%Their SBCR is then still consistent with the one of \citet{Pietr} despite the high arbitrary extinction.}

The contribution of the visual extinction to the SBCR is therefore minimal. However, we decided to take into consideration the extinction since the colour validity interval of the relation can be impacted.

\section{Determination of new surface brightness--colour relations}\label{new_SBCRs}

\subsection{Final selected measurements samples}

With the methodology described in Sect. \ref{Selection}, we obtain four samples of carefully selected measurements, depending on luminosity classes. All the tables (including selected and rejected stars) are provided online. The four tables have the following numbers of selected stars (selected/total\footnote{The total number of measurements is the number of measurements remaining after applying common criteria to the JMDC.}): F5/K7-II/III (70/274), F5/K7-IV/V (38/156), M-II/III (29/67), M-V (16/37). As an example,  the F5/K7 giants sample is shown in Table \ref{sample_table}, including keywords relative to the source of the infrared photometry, as well as specific keywords corresponding to criteria of selection indicated in the "Notes" column. Final selected measurements are those with an empty cell in the Notes column.

\subsection{New specific surface brightness--colour relations}\label{SBCRs_final}

\begin{figure*}
   \centering
   \includegraphics[width=0.5\hsize]{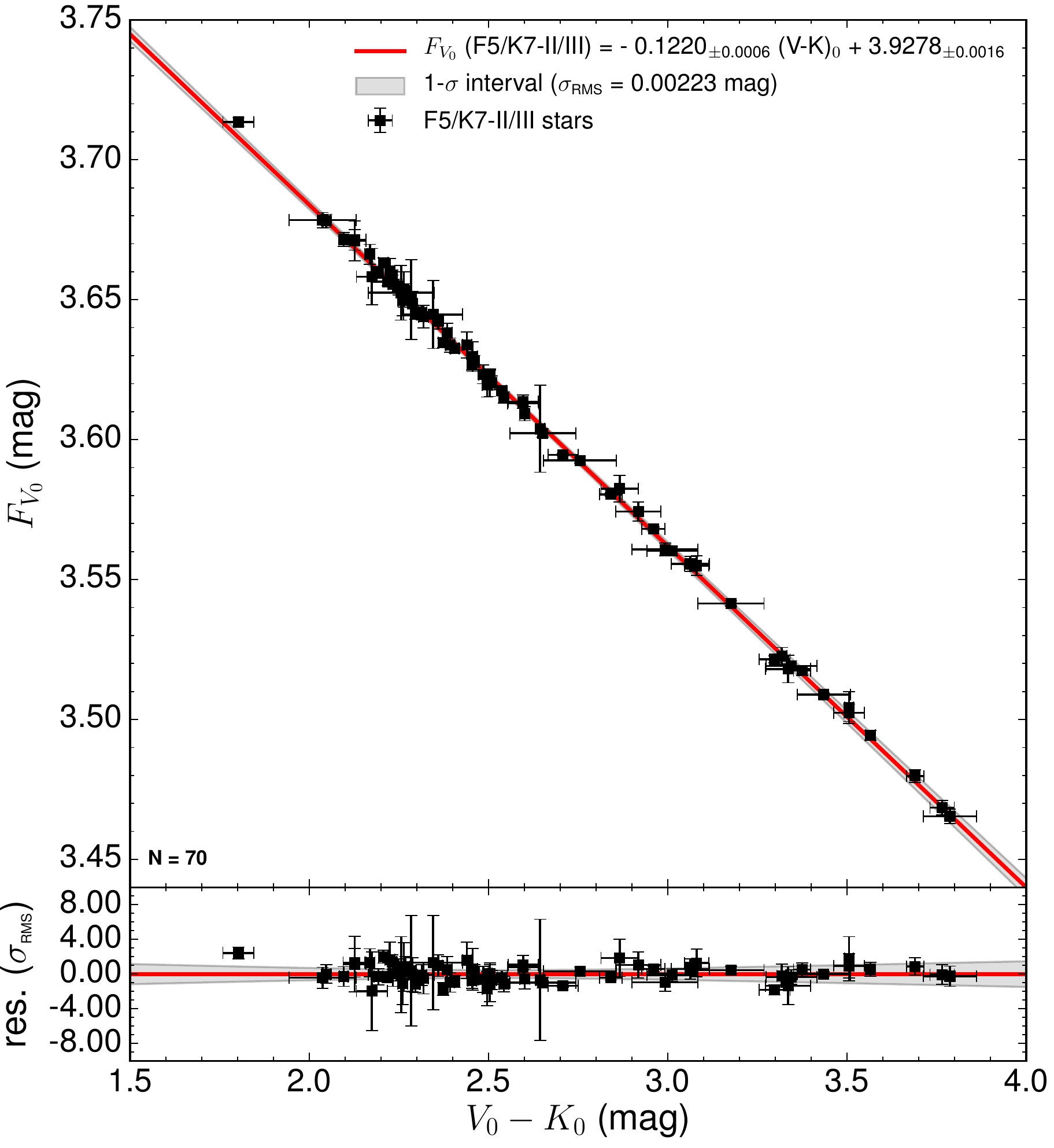}\includegraphics[width=0.50\hsize]{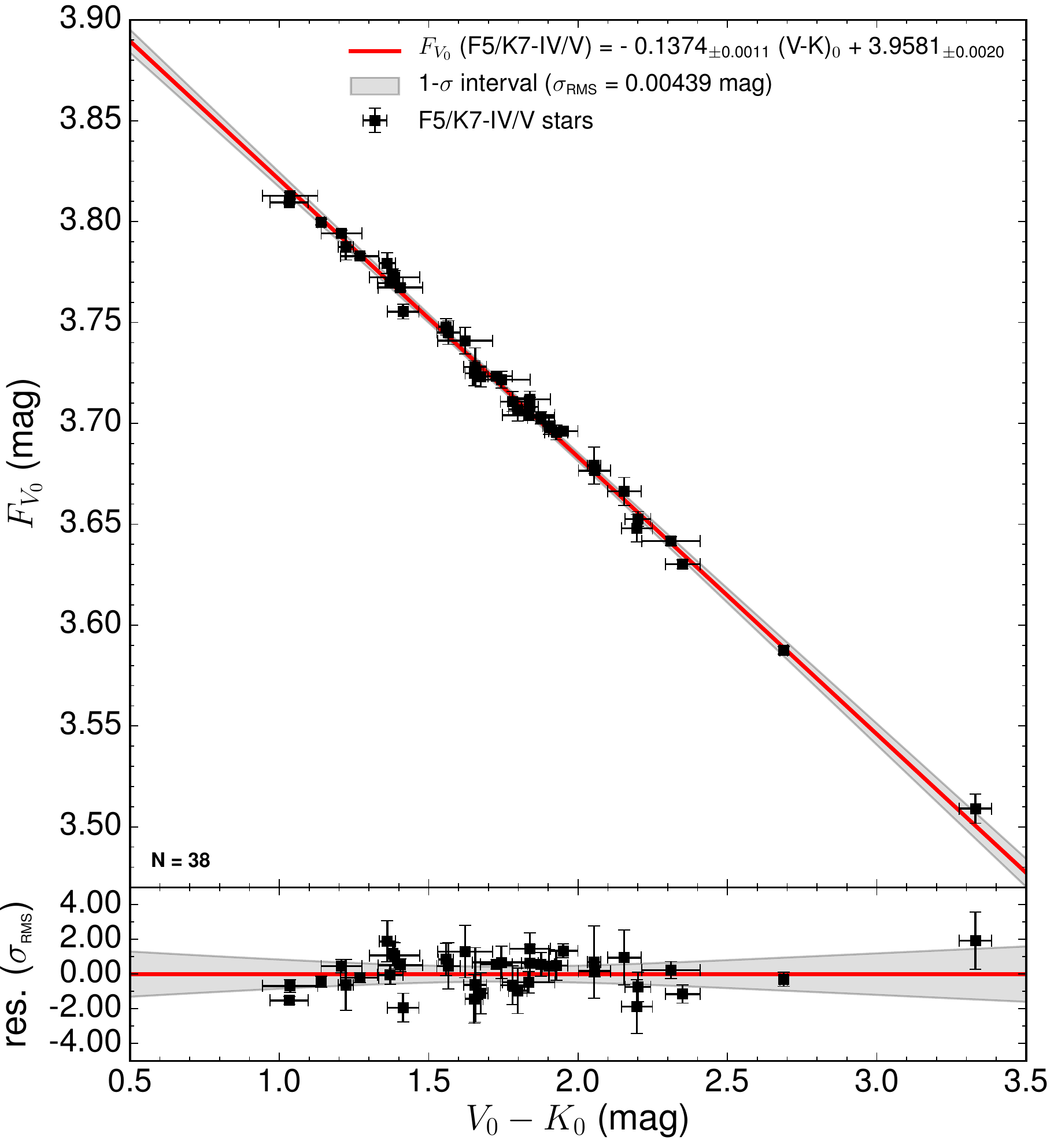}
   \includegraphics[width=0.50\hsize]{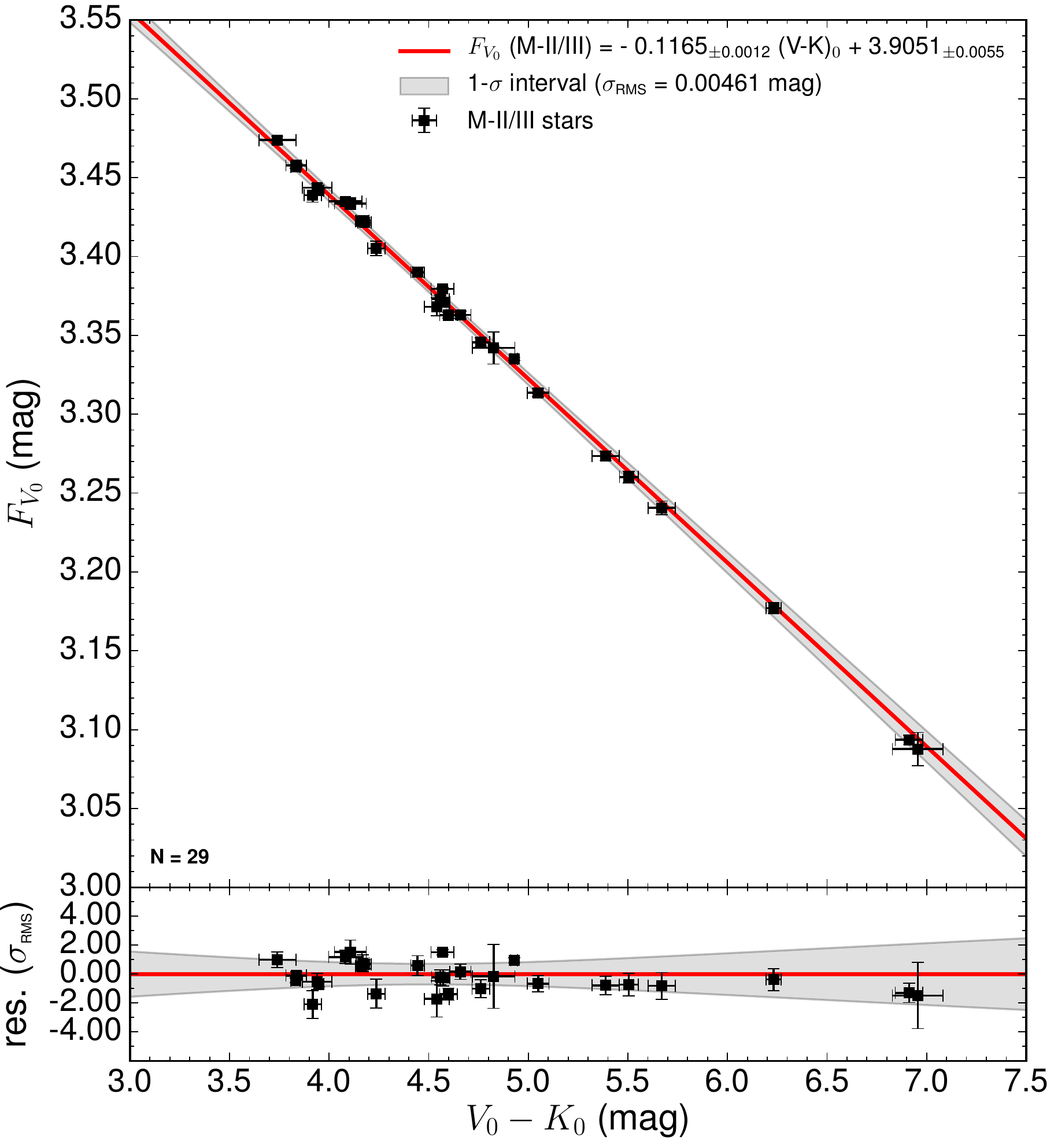}\includegraphics[width=0.50\hsize]{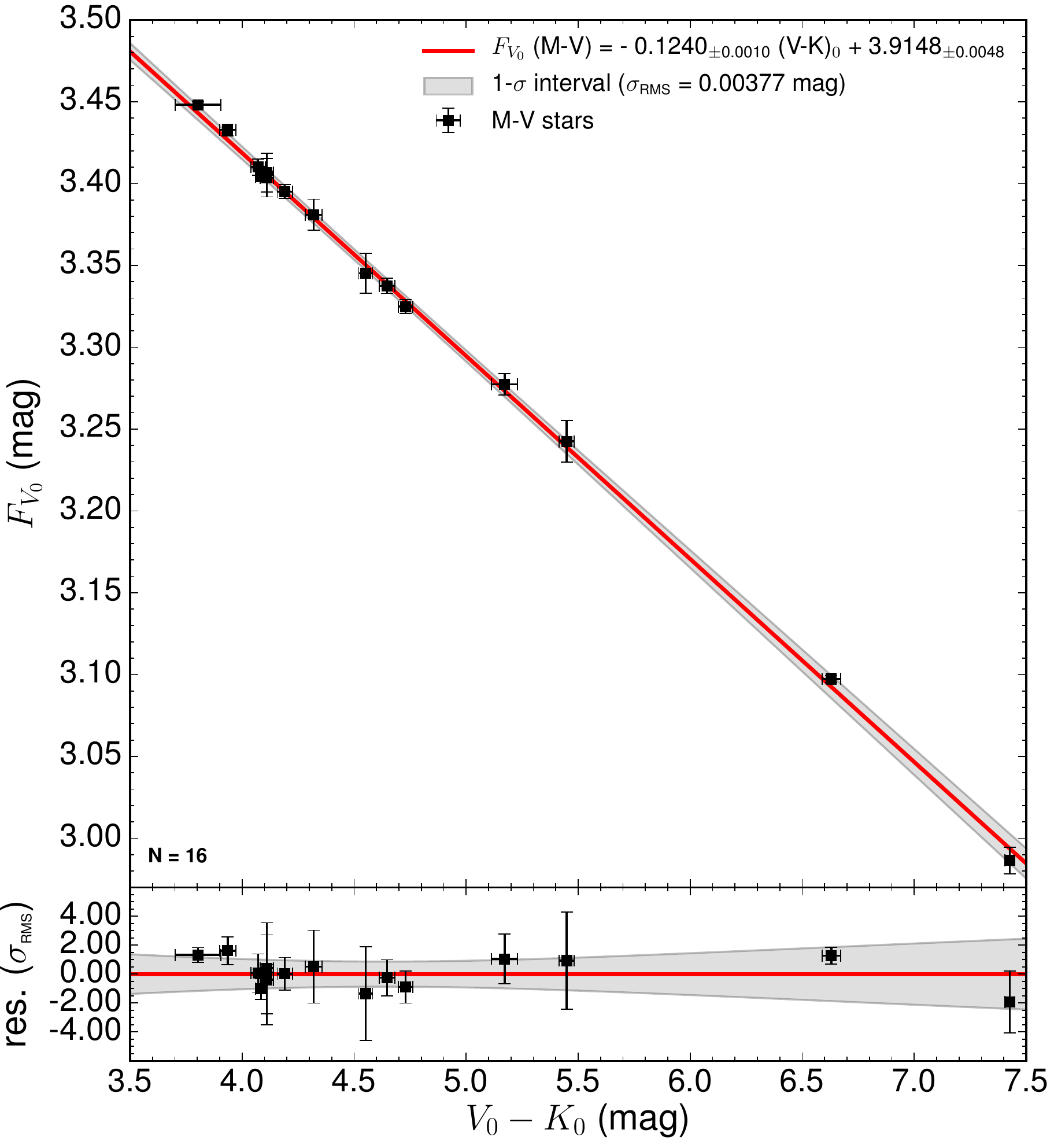}
      \caption{Newly developed surface brightness--colour relations after applying stellar characteristics, interferometric and photometric criteria. From top-left to bottom-right panel: SBCRs for F5/K7 giants, F5/K7 sub-giants/dwarfs, M giants and M dwarfs. The shaded grey area corresponds to the 1-$\sigma$ confidence interval computed according to Eq. \ref{uncertainty_rho}. The uncertainty on the surface brightness of each measurement was divided by the RMS.}
         \label{SBCR_FGKV}
\end{figure*}

\begin{figure}
   \centering
   \includegraphics[width=\hsize]{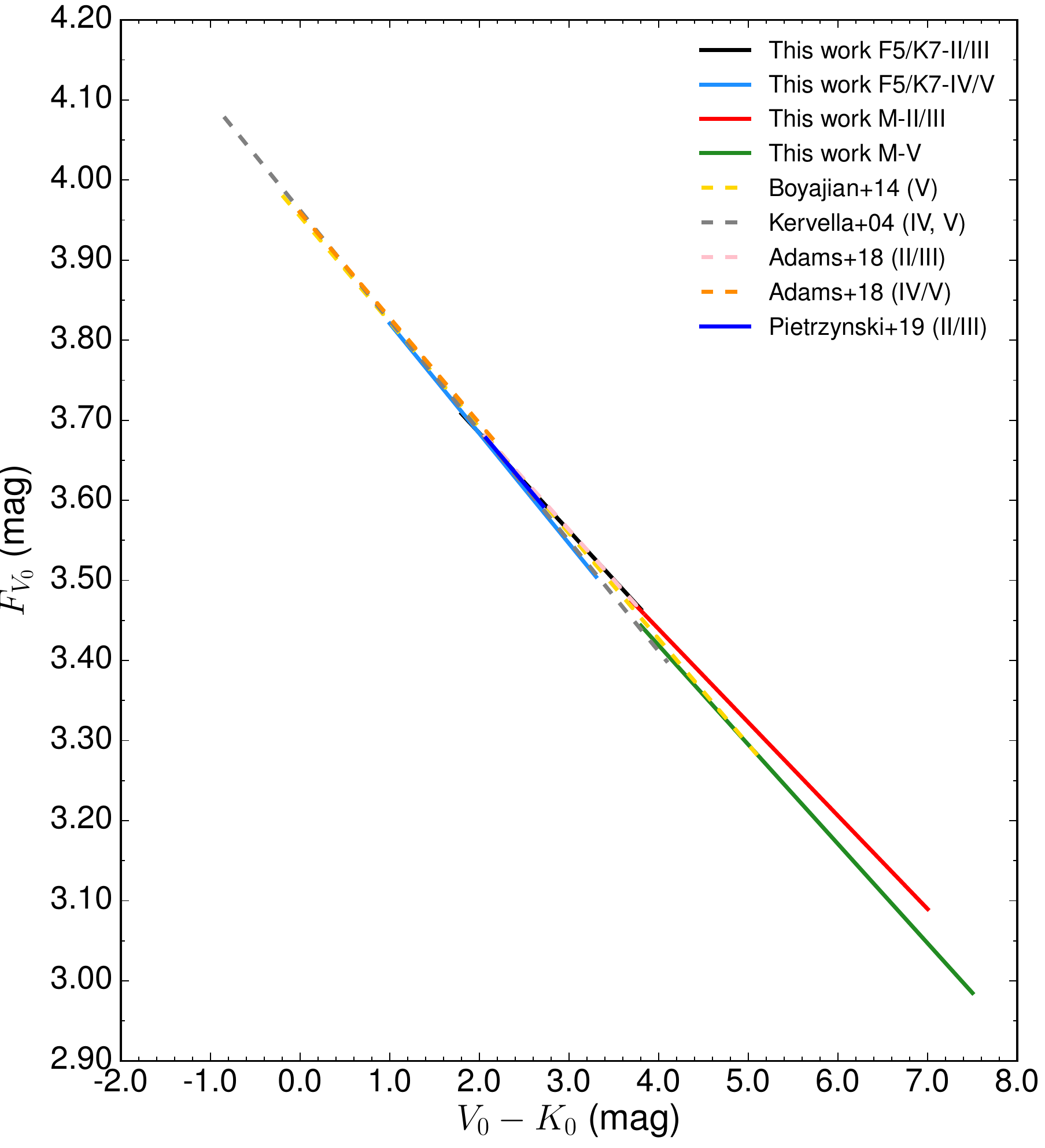}
      \caption{Comparison between our four newly developed SBCRs with relations in the literature.}\label{comparison_SBCRs}
\end{figure}

\begin{figure}
   \centering
   \includegraphics[width=\hsize]{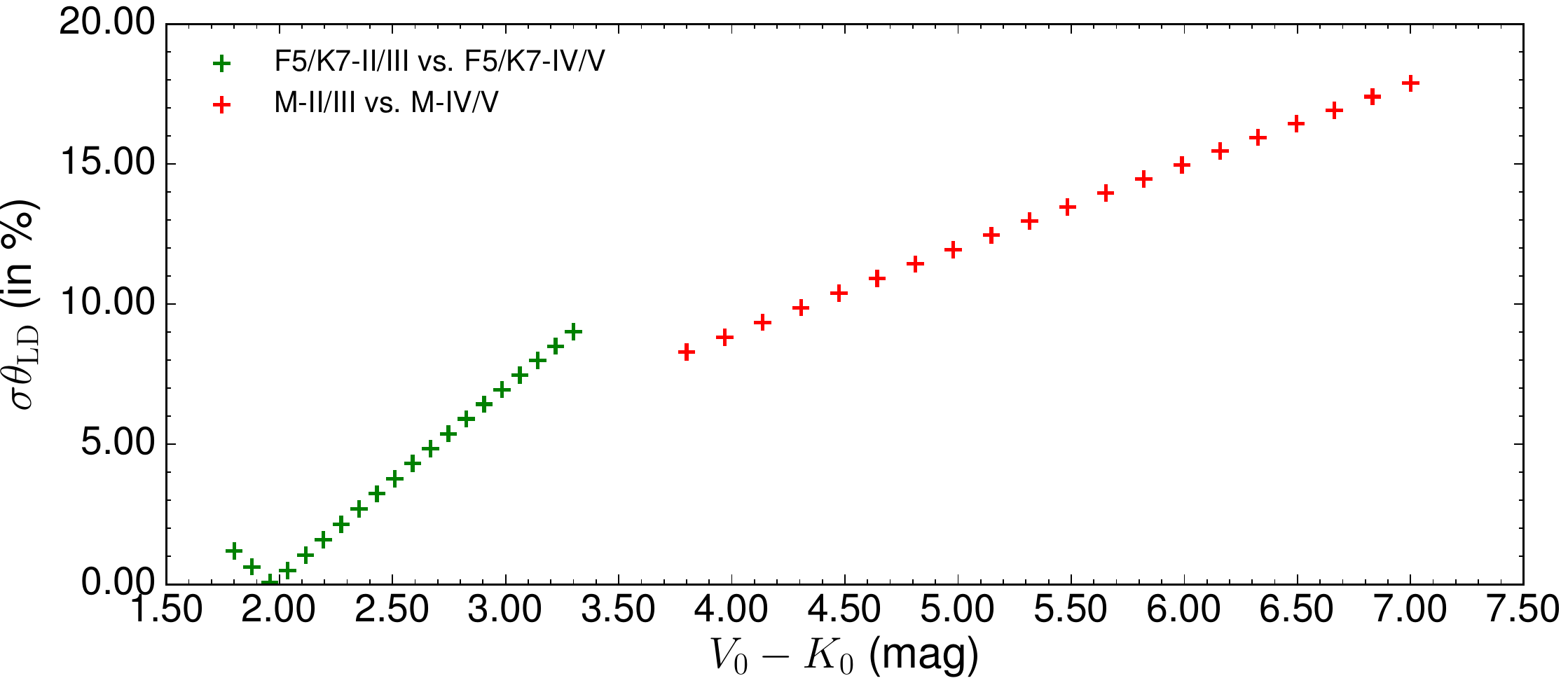}
      \caption{Difference (in \%) of angular diameter estimations between our SBCRs.}\label{comparison_our_SBCRs}
\end{figure}

\begin{figure}
   \centering
   \includegraphics[width=\hsize]{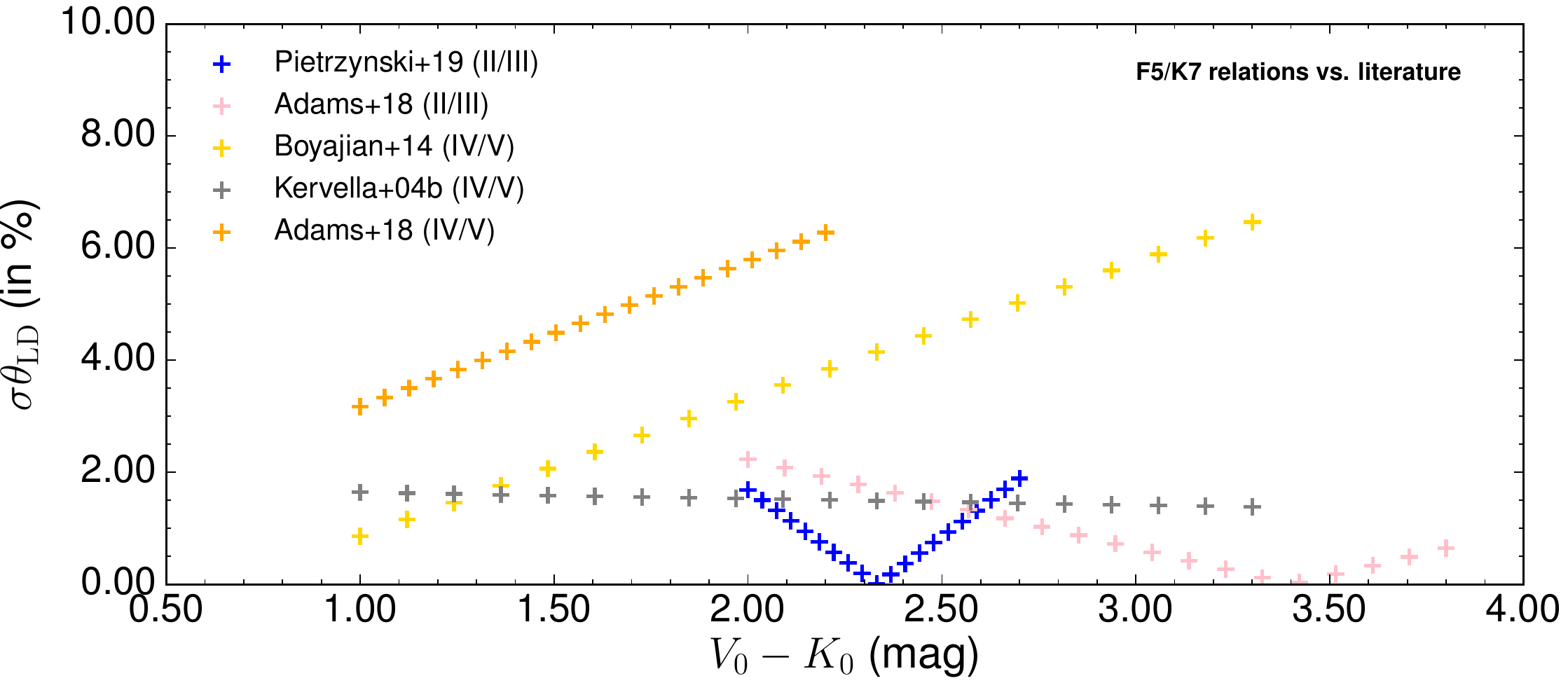}
      \caption{Difference (in \%) of angular diameter estimations between our F5/K7 relations and the literature.}\label{comparison_F_literature}
\end{figure}

\begin{figure}
   \centering
   \includegraphics[width=\hsize]{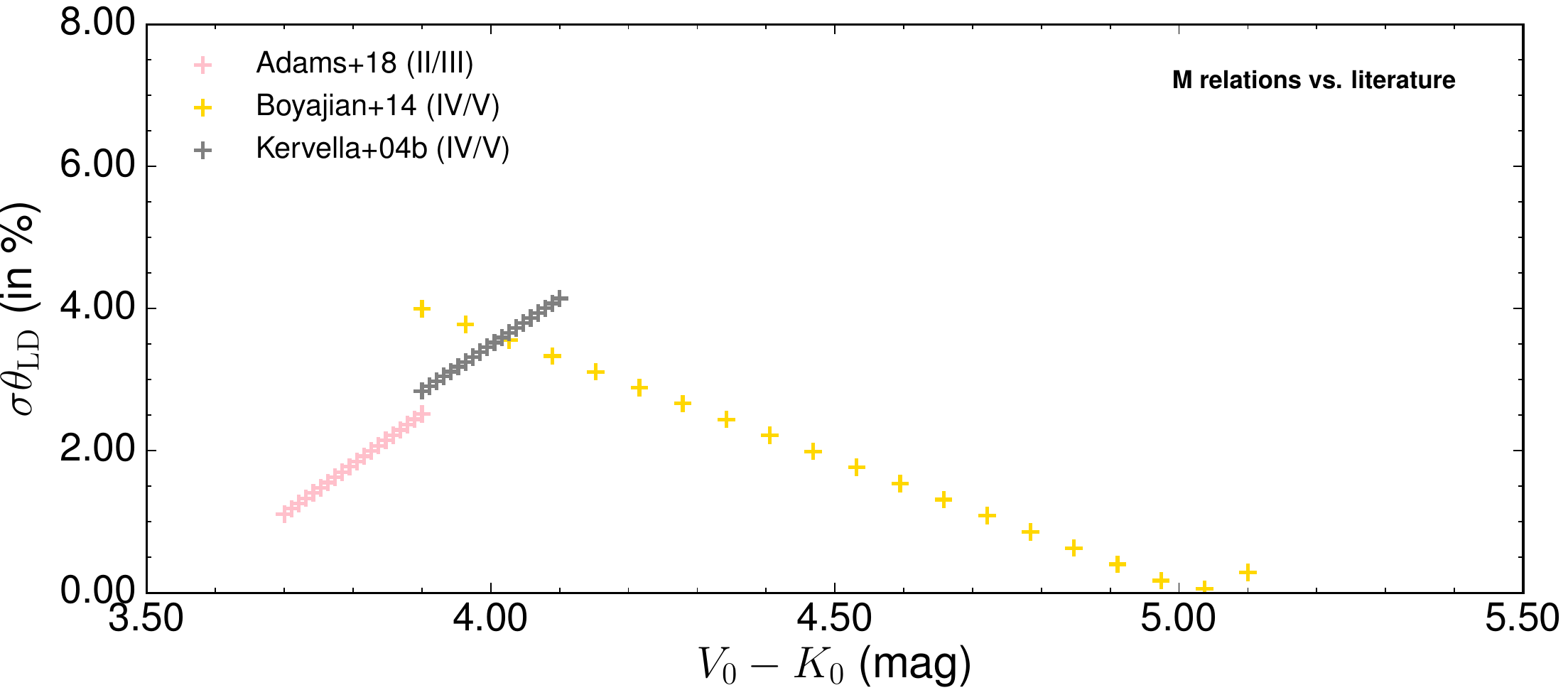}
      \caption{Difference (in \%) of angular diameter estimations between our M relations and the literature.}
         \label{comparison_M_literature}
\end{figure}

\begin{table*}
\caption{Parameters of the new SBCRs. The $(V-K)$ range column denotes the validity interval of the relation.}           
\label{table_final}   
\centering                          % used for centering table
\begin{tabular}{cccccc}        % centered columns (4 columns)
\hline\hline                 % inserts double horizontal lines
Working box & Number of data & Relation & $(V-K)$ range & $\sigma_{\mathrm{RMS}}$ &  Expected $\frac{\sigma_{\theta_{\mathrm{LD}}}}{\theta_{\mathrm{LD}}}$ \\    % table heading 
 & & & [mag] & [mag] & [\%] \\
\hline                        % inserts single horizontal line
F5/K7-II/III & 70 & $F_{V_{0}} = -0.1220_{\pm 0.0006} (V-K)_0 + 3.9278_{\pm 0.0016}$ & [1.80; 3.80] & 0.00223 & 1.03 \\
F5/K7-IV/V & 38 & $F_{V_{0}} = -0.1374_{\pm 0.0011} (V-K)_0 + 3.9581_{\pm 0.0020}$ & [1.00; 3.30] & 0.00439 & 2.02 \\
M-II/III & 29 & $F_{V_{0}} = -0.1165_{\pm 0.0012} (V-K)_0 + 3.9051_{\pm 0.0055}$ & [3.70; 7.00] & 0.00461 & 2.12 \\
M-V & 16 & $F_{V_{0}} = -0.1240_{\pm 0.0010}(V-K)_0 + 3.9148_{\pm 0.0048}$ & [3.80; 7.50] & 0.00377 & 1.73 \\
\hline                                  
\end{tabular}
\end{table*}                                                                    

The new relations for the four samples are presented in Fig. \ref{SBCR_FGKV}. The SBCRs are listed in Table \ref{table_final}, and we detail our fitting strategy in Appendix \ref{fitting_strategy}. We did a test by comparing a least-square (LS) regression with our strategy. We find that using a simple LS method leads to a maximum difference of 1\% on the angular diameter compared to our method. We therefore decided to keep our fitting strategy, since the difference with the LS method is not significant. We consider our method as more robust as we take into consideration all uncertainties that could induce a bias in the final SBCR.

The most precise relation is found for the F5/K7 giants working box, with an RMS of 0.00223 mag. The resulting angular diameter is obtained from Eq. \ref{new_def_SBCR} as follows:

\begin{equation}\label{equation_theta}
    \theta_{\mathrm{LD}} = 10^{8.4392-0.2 V_0-2 F_{V_{0}}}.
\end{equation}

A formal way to calculate the expected angular diameter precision $\sigma_{\theta_{\mathrm{LD}}}$ is to apply the partial derivative method on Eq. \ref{equation_theta}:

\begin{equation}\label{accuracy_equation}
\frac{\sigma_{\theta_{\mathrm{LD}_\mathrm{RMS}}}}{\theta_{\mathrm{LD}}} = 2 \ln(10) \sigma_{\mathrm{RMS}}.
\end{equation}

This leads to a precision of 1\% on the estimation of the angular diameter in the case of F5/K7 giants. 
%Such a precision is comparable to the one recently found by \citet{Pietr}. 
Regarding the other boxes, the RMS of the relations range from 0.00377 mag to 0.00461 mag, leading to an estimate of the angular diameter precision between 1.7\% and 2.1\%. As shown in Table \ref{table_final}, the $V-K$ colour domain of validity of these relations ranges from 1 to 7.5 mag.

However, one should notice that such precision corresponds to a lower limit on the expected angular diameter uncertainty. Indeed, if we want to deduce the angular diameter using a SBCR, we have to consider the uncertainties on the colour and the coefficients of the SBCR. The total resulting uncertainty on the angular diameter can be expressed as $\theta_{\mathrm{LD}}\pm \sigma_{\theta_{\mathrm{LD}_\mathrm{RMS}}} \pm \sigma_{\theta_{\mathrm{LD}_\mathrm{a,b,phot}}}$, where $\sigma_{\theta_{\mathrm{LD}_\mathrm{a,b,phot}}}$ is given by

\begin{equation}
\sigma_{\theta_{\mathrm{LD}_\mathrm{a,b,phot}}} = 2 \ln(10)\theta_{\mathrm{LD}}\sigma_{\theta_{\mathrm{LD}_\mathrm{a,b}}} \sigma_{\theta_{\mathrm{LD}_\mathrm{phot}}},
\end{equation}

where

\begin{equation}
\sigma_{\theta_{\mathrm{LD}_\mathrm{a,b}}} = \left\{\left[(V-K)-0.881A_V\right]^2 \sigma_a ^2 + \sigma_b ^2\right\}^{1/2}
\end{equation}

is the uncertainty linked to the coefficients $a$ and $b$ of the relation, and

\begin{equation}
\sigma_{\theta_{\mathrm{LD}_\mathrm{phot}}} = \left\{a^2 \left(\sigma_V^2 + \sigma_K ^2 + 0.014\sigma_{A_{V}}^2\right)\right\}^{1/2}
\end{equation}

is the photometric part of the uncertainty. For the F5/K7 giants' relation, by considering only uncertainties due to the coefficients of the SBCR, and fixing an arbitrary colour of $V-K = 3$ mag, we find a precision of 1.10\% on the angular diameter. On the other hand, if we consider only arbitrary uncertainties of 0.022 mag on both $V$ and $K$ magnitudes, we get 1.70\% precision. If we set $\sigma_V$ 10\% smaller, we are now only sensitive to $\sigma_K$ and we get 1.20\% precision. This means that precise $V$ and $K$ band photometries ($< 0.022$ mag) are necessary if we want to reach 1\% precision on the angular diameter using an SBCR with a RMS of 0.00223 mag.

A number of interferometric measurements have uncertainties below 1\%. We did a test by setting a lower limit of 1\% on the angular diameter and 0.03 mag on the $V-K$ colour of the stars. The SBCRs we obtained were consistent at less than 1-$\sigma$ with the current ones. A possible under-estimation of the uncertainties therefore has no impact on our SBCRs.

\section{Discussion}\label{discussion}

\subsection{Different surface brightness--colour relations for giants and dwarfs, and a comparison with the literature}\label{discussion_1}

In Fig. \ref{comparison_SBCRs}, we superimposed our SBCRs with various relations found in the literature, namely \citet{2004AA...426..297K}, \citet{Boyajian14}, \citet{Pietr}, and \citet{Adams2018}. In Fig. \ref{comparison_our_SBCRs}, we compare our own relations for giant and dwarf stars, respectively. This shows that using the F5/K7 relation for giants, instead of the one for dwarfs, leads to an error on the estimation of the angular diameter of up to 9\%. The disagreement can even reach 18\% for the M relations. Using relations adapted to the spectral type and class of the star is therefore mandatory. This result is consistent with several previous studies \citep{dB93, Fouque1997, G04,2004AA...426..297K}. 

As mentioned in Sect. \ref{criteria_section}, we decided to ignore the variability criterion for F5/K7 giants. After a case-by-case analysis, we found in \citet{variability} that the variability generates a noise on the $V$ magnitude between $\pm 0.02$ and $\pm 0.10$ mag, with a median value at 0.04 mag. Removing variables from the sample leads to a relation in very good agreement at a level of 0.2$\sigma$ with the current one, but keeping variables does not influence the calibration of our SBCRs.

Recently, \citet{Adams2018} considered 78 giants, sub-giants, and dwarfs, with interferometric angular diameter estimates at the 2\% level or better (and observed on at least two separated occasions), in order to constrain the SBCRs. They used different coulours and a definition of the SBCR compared to the one we use in this work (including $V$ and $K$), and they paid attention to binarity, following the selection strategy described in \citet{2008ApJ...683..424B}. They reached the conclusion (conversely to other authors mentioned above) that there is no difference between the SBCRs of giants, sub-giants, and dwarfs, and they obtained a precision of 3\% in the $V-K$ colour system. Figures \ref{comparison_F_literature} and \ref{comparison_M_literature} show the normalised difference (in \%) on the angular diameter we expect between our SBCRs and relations taken from the literature, introduced above. For dwarfs and sub-giants, we obtained different results to \citet{Adams2018} on the derived angular diameters of at most 6\% over their domain of validity. For F5/K7 and M giants, we respectively obtained a good agreement at the 1.5\% and 2.5\% levels.

\citet{Chelli2016} developed a new method for the calibration of the SBCR based on the differential surface brightness (DSB) and pseudo-magnitudes. Similarly to \citet{Adams2018}, they found a unique polynomial solution for all stars (dwarfs and giants), as a function of the spectral type. This relation gives a precision of about 3\% on the derived angular diameters. If we apply the same methodology of DSB and pseudo-magnitudes on our samples, we obtain a comparable precision to the one obtained for our SBCRs, and, importantly, we again retrieve different DSB relations between giants and dwarfs.

The precision we reached with the F5/K7 giants' SBCR is comparable to the one of \citet{Pietr}. As shown in Fig. \ref{comparison_F_literature}, we expect a difference of at most 2\% on the angular diameter with the SBCR of \citet{Pietr}. This difference could be due to the fact that we considered observational and stellar characteristics selection criteria. We indeed rejected 21 stars among 48 observed by \citet{Pietr} because of their activity, despite the very good quality of the observations. The agreement between the relations is lower than 1.5\% for the majority of the colour range considered. On the other hand, we find very good agreement between our F5/K7-IV/V SBCR and that of \citet{2004AA...426..297K}. Using one or the other relation leads to a difference of less than 1\% on the angular diameter, which reveals a strong consistency of these two SBCRs. However, our M-V relation is inconsistent with the SBCR of \citet{2004AA...426..297K} at a level of more than 4\%, but consistent with \citet{Boyajian14} under 2.5\%. We need new data and complementary works to understand these differences.

\subsection{Surface brightness--colour relations for Gaia}

\begin{figure*}
   \centering
   \includegraphics[width=0.5\hsize]{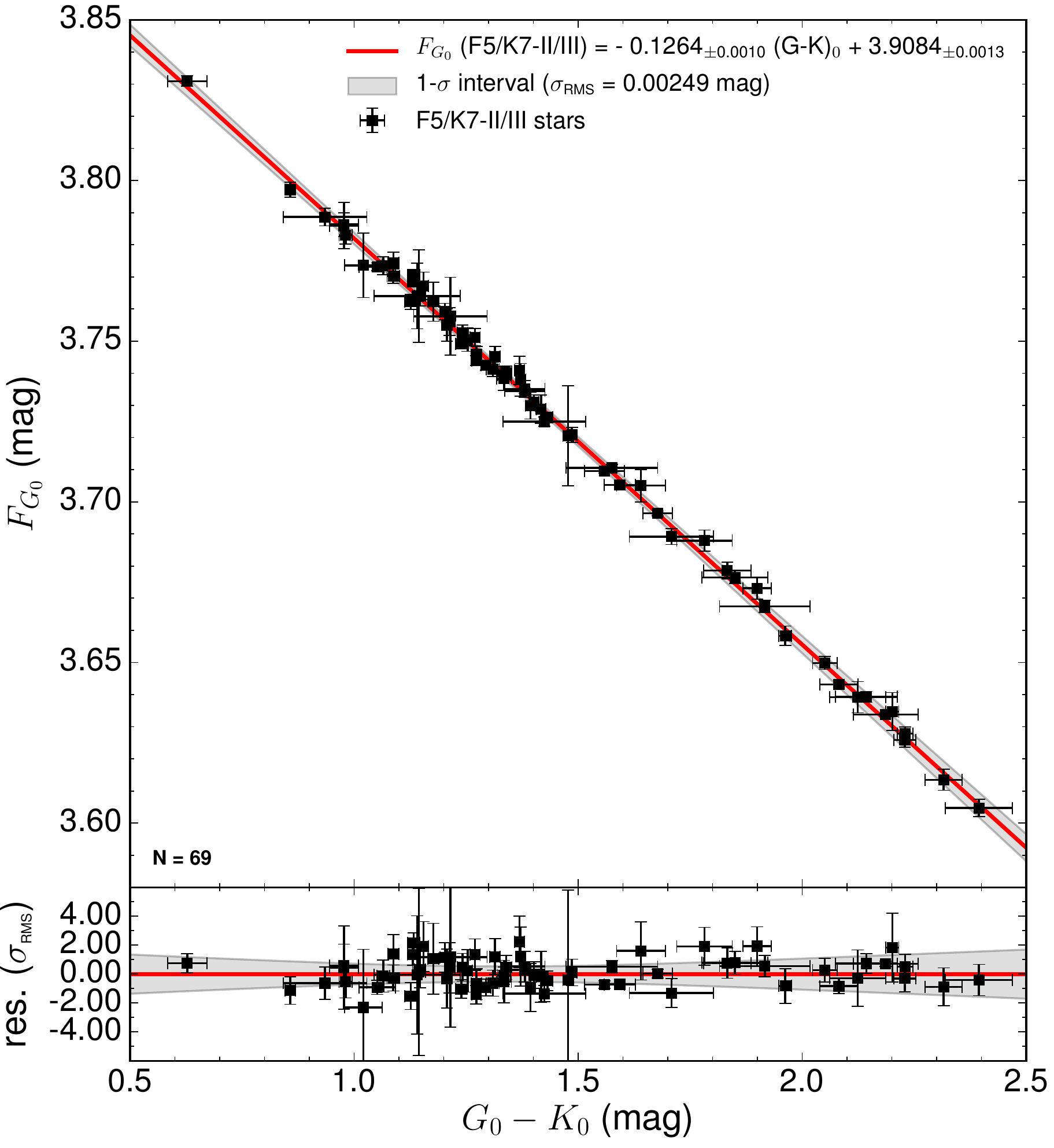}\includegraphics[width=0.50\hsize]{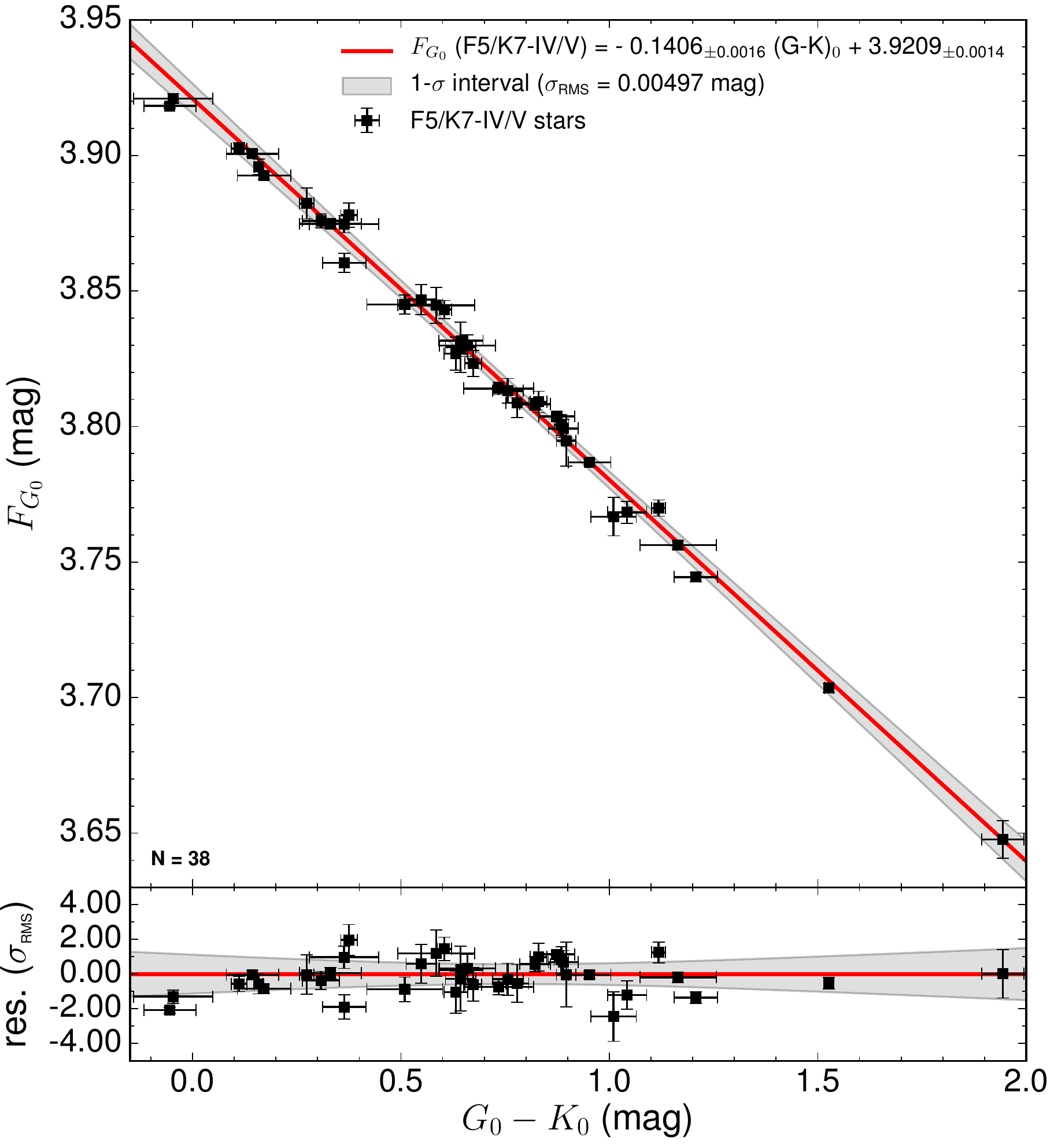}
   \includegraphics[width=0.50\hsize]{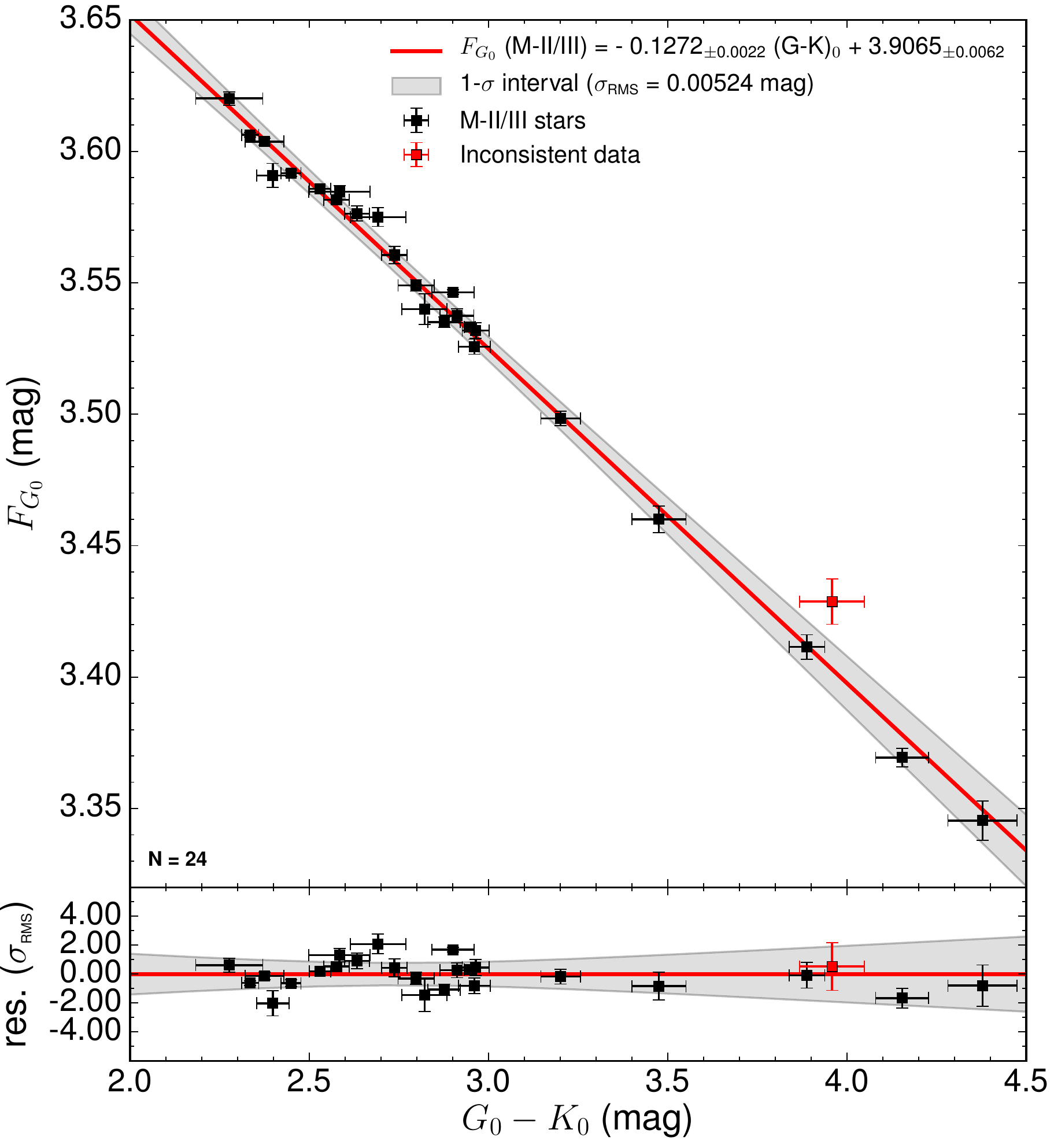}\includegraphics[width=0.50\hsize]{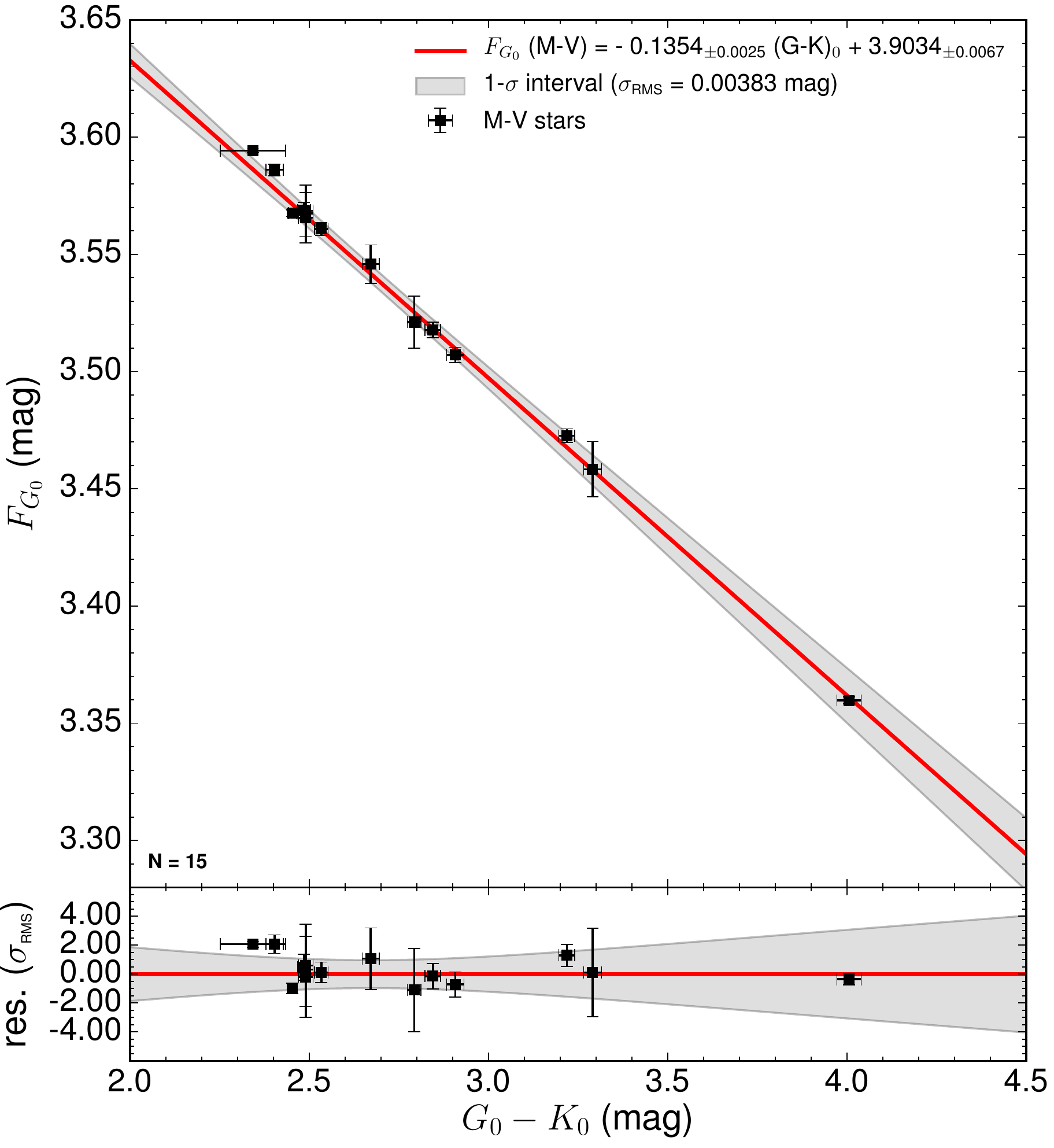}
      \caption{Surface brightness--colour relations based on the ($G$, $G-K$) photometric system. From top-left to bottom-right panel: SBCRs for F5/K7 giants, F5/K7 sub-giants/dwarfs, M giants, and M dwarfs. The shaded grey area corresponds to the 1-$\sigma$ confidence interval computed according to Eq. \ref{uncertainty_rho}.}
         \label{SBCR_Gaia}
\end{figure*}

In this section, we convert our SBCRs in the Gaia photometric band $G$. The $G$ photometry of the stars in our sample is found in the Gaia DR2 database \citep{GaiaCollab}. In order to determine the corresponding extinction ($A_G$), we used an analytic model established by \citet{Danielski}:

\begin{multline}
%\begin{split}
A_G =  a_1 + a_2 (G-K)_0 + a_3 (G-K)_0^2 + a_4 (G-K)_0^3 + a_5 A_V \\
+ a_6 A_V^2+ a_7 (G-K)_0 A_V,
%\end{split}
\end{multline}

with $a_1 = 0.935556283, a_2 = -0.090722012, a_3 = 0.014422056, a_4 = -0.002659072, a_5 = -0.030029634, a_6 = 0.000607315,$ and $a_7 = 0.002713748$. The SBCRs based on the Gaia photometry are shown in Fig. \ref{SBCR_Gaia}, while their coefficients  are listed in Table \ref{table_Gaia}. We find a good consistency with the SBCR based on the V band. Precision ranges from 1.1\% to 2.4\%. 

There are several things to mention. First, we did not find the $G$ photometry for six of the giant stars. Second, one M-II/III star is totally incompatible with the SBCR, namely HD236459 (red point on the bottom-left panel of Fig. \ref{SBCR_Gaia}). Taking a look at this star, its distance is found to be about 2.3kpc (i.e. much further than the distance of the other giants in our sample), leading to a very high visible extinction of $A_V = 1.80$ mag. This star has been removed for the fit of the SBCR.

\begin{table*}
\caption{Parameters of the new SBCRs considering Gaia photometry. The $(G-K)$ range column denotes the validity interval of the relation.}           
\label{table_Gaia}   
\centering                          % used for centering table
\begin{tabular}{cccccc}        % centered columns (4 columns)
\hline\hline                 % inserts double horizontal lines
Working box & Number of data & Relation & $(G-K)$ range & $\sigma_{\mathrm{RMS}}$ &  Expected $\frac{\sigma_{\theta_{\mathrm{LD}}}}{\theta_{\mathrm{LD}}}$ \\    % table heading 
 & & & [mag] & [mag] & [\%] \\
\hline                        % inserts single horizontal line
F5/K7-II/III & 69 & $F_{G_{0}} = -0.1264_{\pm 0.0010} (G-K)_0 + 3.9084_{\pm 0.0013}$ & [0.50; 2.40] & 0.00249 & 1.15 \\
F5/K7-IV/V & 38 & $F_{G_{0}} = -0.1406_{\pm 0.0016} (G-K)_0 + 3.9209_{\pm 0.0014}$ & [-0.10; 2.00] & 0.00497 & 2.29 \\
M-II/III & 24 & $F_{G_{0}} = -0.1272_{\pm 0.0022} (G-K)_0 + 3.9065_{\pm 0.0062}$ & [2.20; 4.40] & 0.00524 & 2.41 \\
M-V & 15 & $F_{G_{0}} = -0.1354_{\pm 0.0025}(G-K)_0 + 3.9034_{\pm 0.0067}$ & [2.30; 4.00] & 0.00383 & 1.77 \\
\hline                                  
\end{tabular}
\end{table*}

\subsection{Validating our methodology with recent interferometric measurements}\label{VEGA/PAVO}

To go further in the validation of our methodology, we selected 10 new stars for interferometric observations with both the Precision Astronomical Visible Observations (PAVO) \citep{PAVO} and the Visible spEctroGraph and polArimeter (VEGA) \citep{Mourard2009, Mourard2011} instruments. These instruments are installed on the Center for High Angular Resolution Astronomy (CHARA) array, in Mount Wilson, USA \citep{CHARA}. Comparing interferometric measurements from different instruments serves to support the importance of such selection criteria to implement our SBCRs. The ten stars were observed between July 2013 and August 2016 with PAVO, and from August 2012 to June 2019 with VEGA. They have spectral types between G6 and K3. Eight of them are giants, one is a sub-giant and one is a dwarf.  The data analysis and results of VEGA measurements are briefly presented here, while the PAVO measurements, as well as a careful comparison of the VEGA and PAVO data will be presented in a forthcoming separate paper \citep{Orlagh2020}. For the data analysis, we used the standard approach described in \citep{Mourard2009, Mourard2011}. We fitted a limb-darkening model to the VEGA visibility measurements using the LITpro software \citep{LITpro}. The $u_{R}$ linear to limb-darkening coefficient for each star was found using the \citet{Claret} catalogue. Results are shown on the top part of Table \ref{VEGA_PAVO}. The corresponding visibility curves are included in Fig. \ref{vis_fig} of the appendix. In order to complete the analysis, we also added four giant stars, recently observed by CHARA/VEGA and presented in \citet{Nardetto2020}. They compare their limb-darkened angular diameters to the ones derived in the $H$-band with the Precision Integrated Optics Near-infrared Imaging ExpeRiment (PIONIER) \citep{Pionier} on VLTI. Results are listed in the bottom part of Table \ref{VEGA_PAVO}. The important point is that all stars in  Table \ref{VEGA_PAVO} have been observed by two different instruments, and the derived angular diameters are found to be consistent at the 1$\sigma$ level. These limb-darkened angular diameters are thus extremely robust.

\begin{table*}
\caption{Top: our new VEGA angular diameter measurements for ten stars. Bottom: VEGA measurements from \citet{Nardetto2020}. We included a "Notes" column that refers to selection criteria if the star is concerned.}           
\label{VEGA_PAVO}   
\centering                          % used for centering table
\begin{tabular}{cccccccc}        % centered columns (4 columns)
\hline\hline                                                                                                    
Name    &       Sp. Type        &       $A_V$   &       $(V-K)_0$       &       $u_{\mathrm{R}}$        &       $\theta_{\mathrm{LD}}$         &       $\chi^2_{r}$ & Notes    \\
        &               &       [mag]   &       [mag]   &               &       [mas]   & &               \\
\hline
HD167042        &       K1III   &       0       &       $2.415_{\pm  0.242}$    &       0.649   &       $0.831_{\pm  0.068}$        &       2.994 & hK      \\
HD175740        &       G8III   &       0       &       $2.484_{\pm  0.063}$    &       0.651   &       $1.145_{\pm 0.012}$ &       0.635 & M       \\
HD178208        &       K3III   &       0.053   &       $2.869_{\pm  0.282}$    &       0.723   &       $1.071_{\pm  0.011}$        &       0.480 & SB      \\
HD180756        &       G8III   &       0.037   &       $2.025_{\pm  0.040}$    &       0.635   &       $0.683_{\pm  0.016}$        &       0.214 & V \\
HD181069        &       K1III   &       0.062   &       $2.478_{\pm  0.040}$    &       0.692   &       $0.763_{\pm  0.010}$        &       0.511 & V       \\
HD181597        &       K1III   &       0       &       $2.660_{\pm  0.292}$    &       0.695   &       $0.892_{\pm  0.007}$        &       0.274 & hK      \\
HD182896        &       K0III   &       0.053   &       $2.643_{\pm  0.026}$    &       0.707   &       $0.710_{\pm  0.024}$        &       0.375 & V       \\
HD185657        &       G6V     &       0       &       $2.370_{\pm  0.220}$    &       0.641   &       $0.726_{\pm  0.006}$        &       0.401 & SB      \\
HD21467 &       K0IV    &       0       &       $2.433_{\pm  0.222}$    &       0.664   &       $0.881_{\pm  0.011}$        &       0.473 & M \\
HD73665 &       G8III   &       0       &       $2.138_{\pm  0.018}$    &       0.660   &       $0.621_{\pm  0.009}$        &       0.125 & M \\
\hline                                                                                                  
HD13468 &       G9III   &       0.028   &       $2.248_{\pm  0.020}$    &       0.648   &       $0.902_{\pm  0.017}$        &       1.600 & -       \\
HD23526 &       G9III   &       0.053   &       $2.228_{\pm  0.020}$    &       0.652   &       $0.920_{\pm  0.028}$        &       0.600 & -       \\
HD360   &       G8III   &       0.028   &       $2.311_{\pm  0.020}$    &       0.680   &       $0.888_{\pm  0.010}$        &       0.500 & -       \\
HD40020 &       K2III   &       0.040   &       $2.434_{\pm  0.020}$    &       0.690   &       $1.033_{\pm  0.022}$        &       0.400 & -       \\
\hline
\end{tabular}
\end{table*}

\begin{figure*}
   \centering
   \includegraphics[width=0.505\hsize]{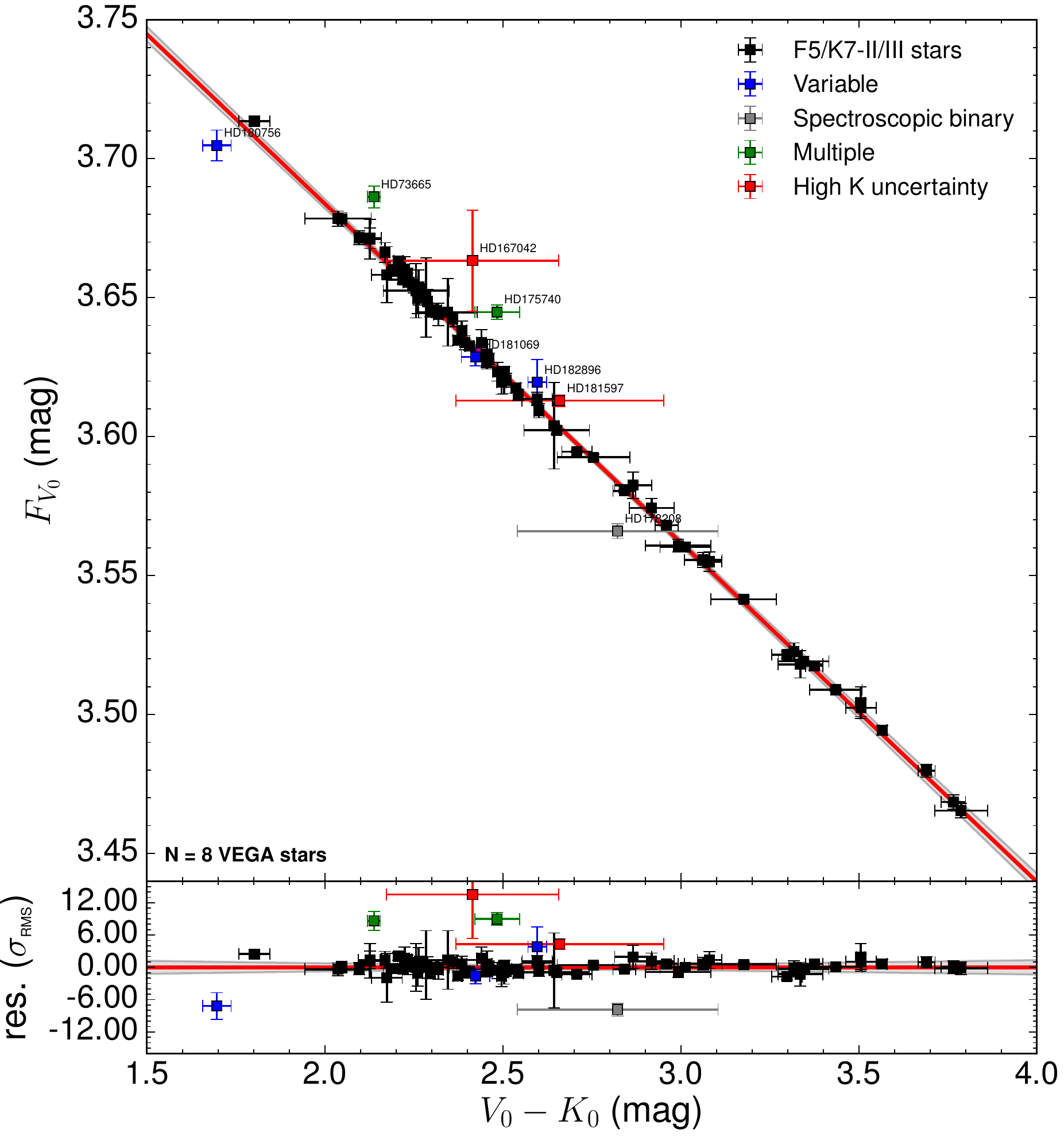}\hfill\includegraphics[width=0.495\hsize]{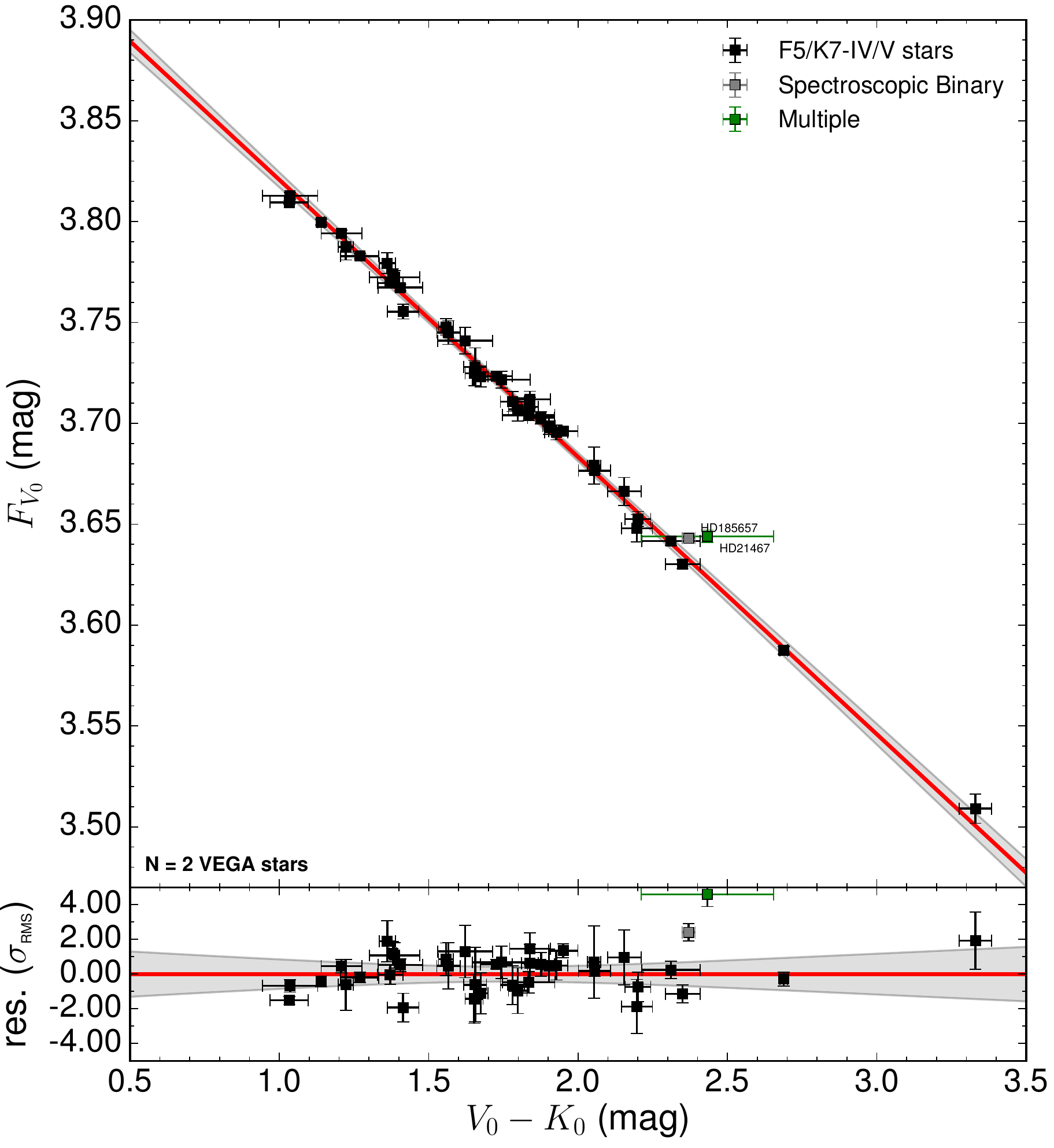}
      \caption{Left: the relation for F5/K7 giant stars including the eight new VEGA measurements. Right: same for F5/K7 sub-giants/dwarfs with two new VEGA measurements.}\label{10more}
\end{figure*}

\begin{figure}
   \centering
   \includegraphics[width=\hsize]{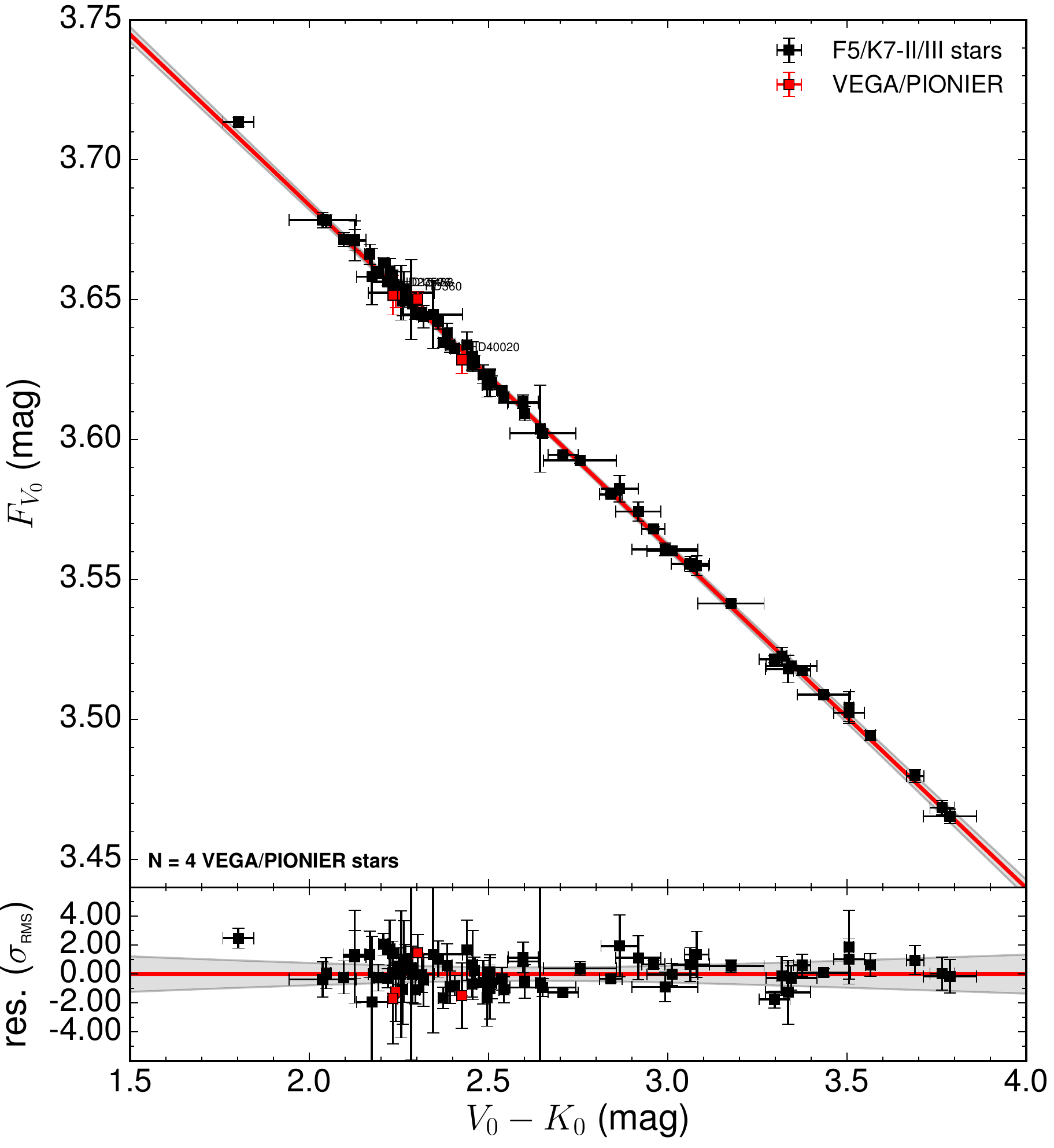}
      \caption{Relation for F5/K7 giant stars including the four VEGA measurements from \citet{Nardetto2020}.}\label{4more}
\end{figure}

Left and right panels of Fig. \ref{10more} show the ten stars observed by VEGA and PAVO on their corresponding SBCR. We find that these stars are not consistent with our relations at a level of up to 12$\sigma$. Looking at the selection criteria described in Sect. \ref{Selection}, these stars should be rejected from the sample, because of multiplicity, variability, and poor $K$ photometry, as indicated in the Notes column of Table \ref{VEGA_PAVO}. This result demonstrates the importance of the selection criteria that we have defined.
The four giant stars observed by VEGA and PIONIER \citep{Nardetto2020} fulfill all the selection criteria. The results are shown in Fig. \ref{4more} above the F5/K7 giants' relation. As expected, the measurements are totally consistent with the SBCR at a level of $\sim$1.5\%. This supports the existence of such selection criteria to constrain SBCRs.

\section{Conclusion and perspectives}

We considered all the interferometric measurements of angular diameters obtained so far in order to build accurate SBCRs. We also refined the methodology by homogeneously applying a list of selection criteria. Combining our new VEGA interferometric measurements with those of \citet{Nardetto2020} and \citet{Orlagh2020}, we demonstrated the coherence of our criteria and the importance they have in the determination of the SBCRs. The variability, the multiplicity, along with other stellar characteristics diagnostics, or even the quality of the interferometric observations, as well as the spatial frequency coverage, appear to be of high importance in building consistent SBCRs. 

Using this approach, we reinforce the conclusion that the surface brightness of a star depends on its spectral type and its luminosity class, since our new SBCRs for giants and sub-giants/dwarfs are inconsistent with each other at a level of up to 18\% on the derived angular diameter, depending on the SBCR considered. Using these criteria, we developed four SBCRs that allow us to estimate angular diameters with an accuracy between 1\% and 2\%, as soon as the precision of the magnitude of the star is better than 0.04 magnitude. 

The objective is to use these SBCRs in the context of PLATO, in order to infer the radii of stars and planets. Our SBCRs were implemented consistently with the PLATO specifications in terms of spectral type and classes. Moreover, our results are consistent in terms of precision with the PLATO objectives since the spatial mission is expected to bring stellar radii measurements with less than 2\% precision. However, our sample of stars still has to be enlarged by an order to magnitude in order to improve the robustness of the SBCRs. Using the Stellar Parameters and Images with a Cophased Array (SPICA) instrument at the focus of the CHARA array, we expect to derive the angular diameter of 800 stars in a few years with a 1\% precision level \citep{SPICA}, which should definitively improve our knowledge of SBCRs. In this context, using the Gaia and 2MASS photometric systems, which both have the largest databases, seems to be the best approach.

Since the Gaia photometry is among the most homogeneous over the full sky, we calibrate SBCRs using this precise photometry for the first time. We reach a precision on the angular diameter between 1.1\% and 2.4\%.

\begin{acknowledgements}
A. S. acknowledges F. Arenou and N. Leclerc for their contribution to the computation of the interstellar extinction with \textit{Stilism}. This work is based upon observations obtained with the Georgia State University Center for High Angular Resolution Astronomy Array at Mount Wilson Observatory. The CHARA Array is funded by the National Science Foundation through NSF grants AST-0606958 and AST-0908253 and by Georgia State University through the College of Arts and Sciences, as well as the W. M. Keck Foundation. This work made use of the JMMC Measured stellar Diameters Catalog \citep{Duvert}. This research made use of the SIMBAD and VIZIER\footnote{Available at \url{http://cdsweb.u- strasbg.fr/}} databases at CDS, Strasbourg (France) and the electronic bibliography maintained by the NASA/ADS system. This work has made use of data from the European Space Agency (ESA) mission Gaia (\url{https://www.cosmos.esa.int/gaia}). This research also made use of Astropy, a community-developed core Python package for Astronomy \citep{astropy2018}.
\end{acknowledgements}

% WARNING
%-------------------------------------------------------------------
% Please note that we have included the references to the file aa.dem in
% order to compile it, but we ask you to:
%
% - use BibTeX with the regular commands:
%   \bibliographystyle{aa} % style aa.bst
%   \bibliography{Yourfile} % your references Yourfile.bib
%
% - join the .bib files when you upload your source files
%-------------------------------------------------------------------

\bibliographystyle{aa}
\bibliography{salsi.bib}

\begin{appendix}

\section{Fitting strategy}\label{fitting_strategy}

In the traditional case, we suppose that only $y$ is subject to measurement error, and $x$ is observed without error. In this work, we built our fitting strategy around the orthogonal distance regression (ODR), which, contrary to the ordinary least-squares (OLS) regression, considers both $x$ and $y$ errors, respectively errors on the $V-K$ colour and on the surface brightness $F_V$ in our case. If we consider that all variables $x_i$ and $y_i$ are respectively affected by the errors $\delta _i \in \mathbb{R}$ and $\epsilon _i \in \mathbb{R}$, the representative model is then written as

\begin{equation}
y_i = f(x_i+\delta _i ; \beta) - \epsilon _i, ~~~i \in 1;...;N ,
\end{equation}

where $\beta$ are the parameters of the fitted model, and $N$ the number of measurements. The ODR method consists of finding the parameter $\beta$ that minimises the sum of orthogonal distances (that we label as $r$ here) between the data points and the fit. The condition to be respected is

\begin{equation}
r = \underset{\beta, \delta, \epsilon}{\min}  \frac{1}{2} \sum_{i=1}^{N} \left(\omega _{\delta_{i}} \delta _i^2 + \omega _{\epsilon_{i}} \epsilon _i ^2 \right),
\end{equation}

where $\omega _i = 1/\sigma_i ^2$ is a weighting, introduced to compensate for instance when $y_i$ and $x_i$ have unequal precision. The final accuracy of the relation is deduced from the RMS $\sigma_{\mathrm{RMS}}$, that we compute in the following way:

\begin{equation}
\sigma_{\mathrm{RMS}} = \sqrt{\frac{1}{N} \times \sum_{i=1}^{N} \left(F_{V_{\mathrm{obs}_i}}-F_{V_{\mathrm{fit}_i}} \right)^2 },
\end{equation}

where $F_{V_{\mathrm{obs}_i}}$ is the measured surface brightness, and $F_{V_{\mathrm{fit}_i}}$ is the fit deduced from the ODR method introduced above. We then wanted to estimate the expected dispersion of surface brightnesses as a function of  the $V-K$ colour of the star, according to our newly developed SBCR. Most authors compute the RMS of the relation and plot a constant expected accuracy around their measurements. However, the weight of the measurements should be taken into account. Indeed, the number of measurements at a given colour indicates how precise the relation is. Following \citet{Gallenne17}, we searched for the barycentre of the measurements. To give clarity to our calculations, we note $C$ as the $V-K$ colour. In this sense, the linear fit is written as follows:

\begin{equation}
F_{V} = a \left(C - C_0\right)+b,
\end{equation}

where $C_0$ is the barycentre of the measurements. The extrapolated uncertainty on the model is then given by

\begin{equation}\label{uncertainty_rho}
\sigma_{F_{V}} = \left(C-C_0\right)^2 \sigma_a ^2 + \sigma_b ^2,
\end{equation}

where $\sigma_a$ and $\sigma_b$ are, respectively, the uncertainties on the coefficient $a$ and $b$ of the fit. The condition to be satisfied here is finding $C_0$ so that $\rho = \partial r^2 / \partial a \partial b$ is zero, where $r^2$ is the distance between the fit and the data. The coefficient $\rho$ corresponds to the correlation between $a$ and $b$. With the previous condition, Eq. \ref{uncertainty_rho} becomes

\begin{equation}
\sigma_{F_{V}} = \sqrt{C^2 \sigma _a ^2 + \sigma _b^2 + 2 \rho \sigma _a  \sigma _b C}.
\end{equation}

For any dataset ($F_{V_{i}} \pm \sigma_i, C_i)$ and any value of $C_0$,  the correlation $\rho$ between $a$ and $b$ can be expressed as

\begin{equation}\label{rho_complicated}
\rho = -\frac{\partial^2 r^2/\partial a \partial b}{\sqrt{\left(\partial^2 r^2 / \partial a^2\right)\left(\partial ^2 r^2 / \partial b^2\right)}} = - \frac{\sum_{i} \frac{C_i-C_0}{\sigma_{i}^2}}{\sqrt{\sum_{i} \frac{\left(C_i-C_0\right)^2}{\sigma_{i}^2}\sum_{i} \frac{1}{\sigma_{i}^2}}} .
\end{equation}

In order to simplify this equation into a more convenient form for estimating an order of magnitude, we made some basic assumptions. First, all the uncertainties $\sigma_i$ on the surface brightness are equal, and then we assume $C_0 = 0$. In this particular case, Eq. \ref{rho_complicated} becomes

\begin{equation}
\rho = -\frac{ \sum_i C_i}{\sqrt{N \sum_i C_i^2}}.
\end{equation}

This method has the advantage of carefully estimating the extrapolated uncertainty of the linear model, depending on the number of measurements made at a given colour. This indicates that our SBCR will be more precise around the $V-K$ colour where most of the measurements were done. 

\section{VEGA visibility curves}
\setcounter{figure}{1}  
\begin{figure*}
   \centering
   \includegraphics[width=0.5\hsize]{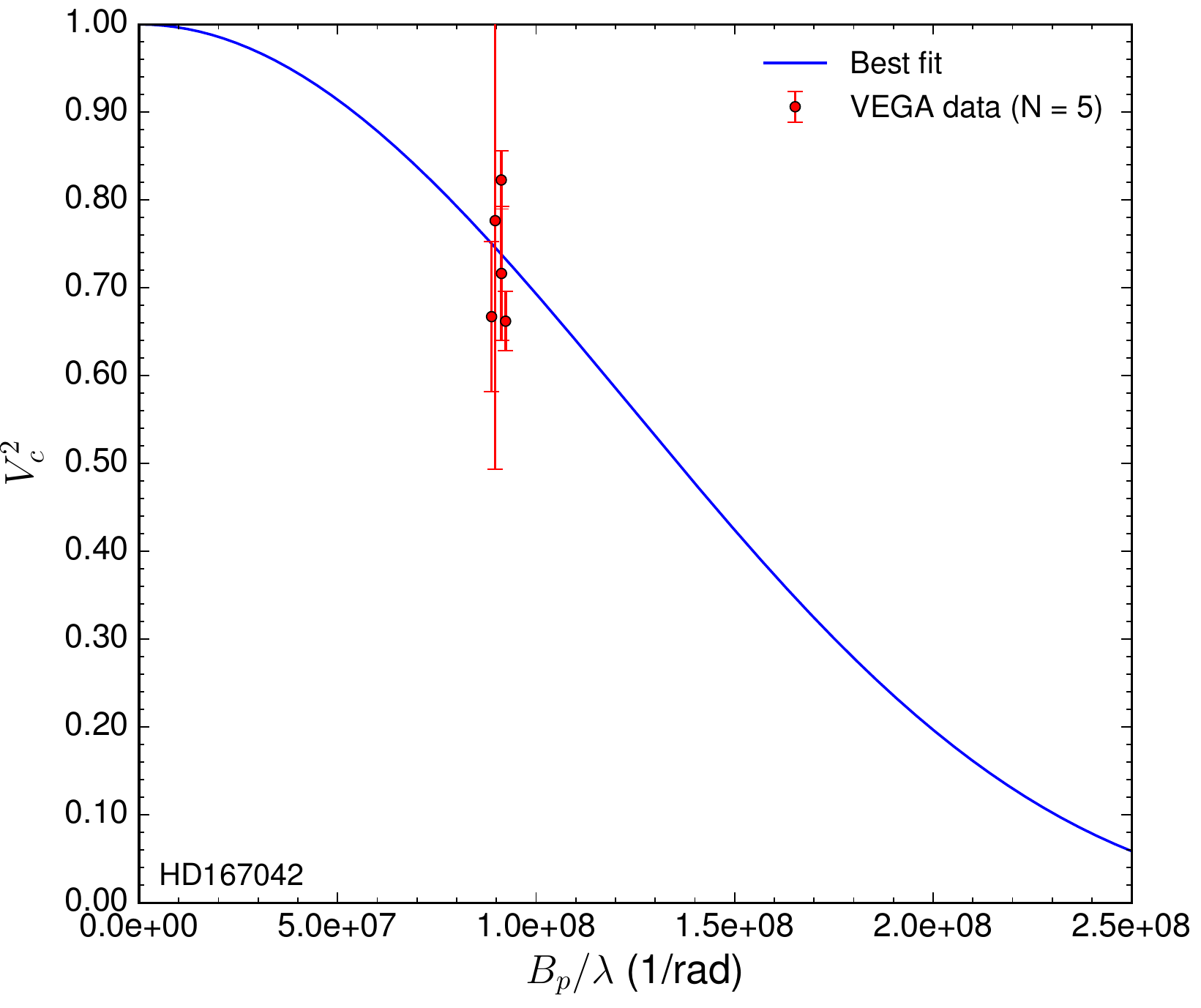}\includegraphics[width=0.5\hsize]{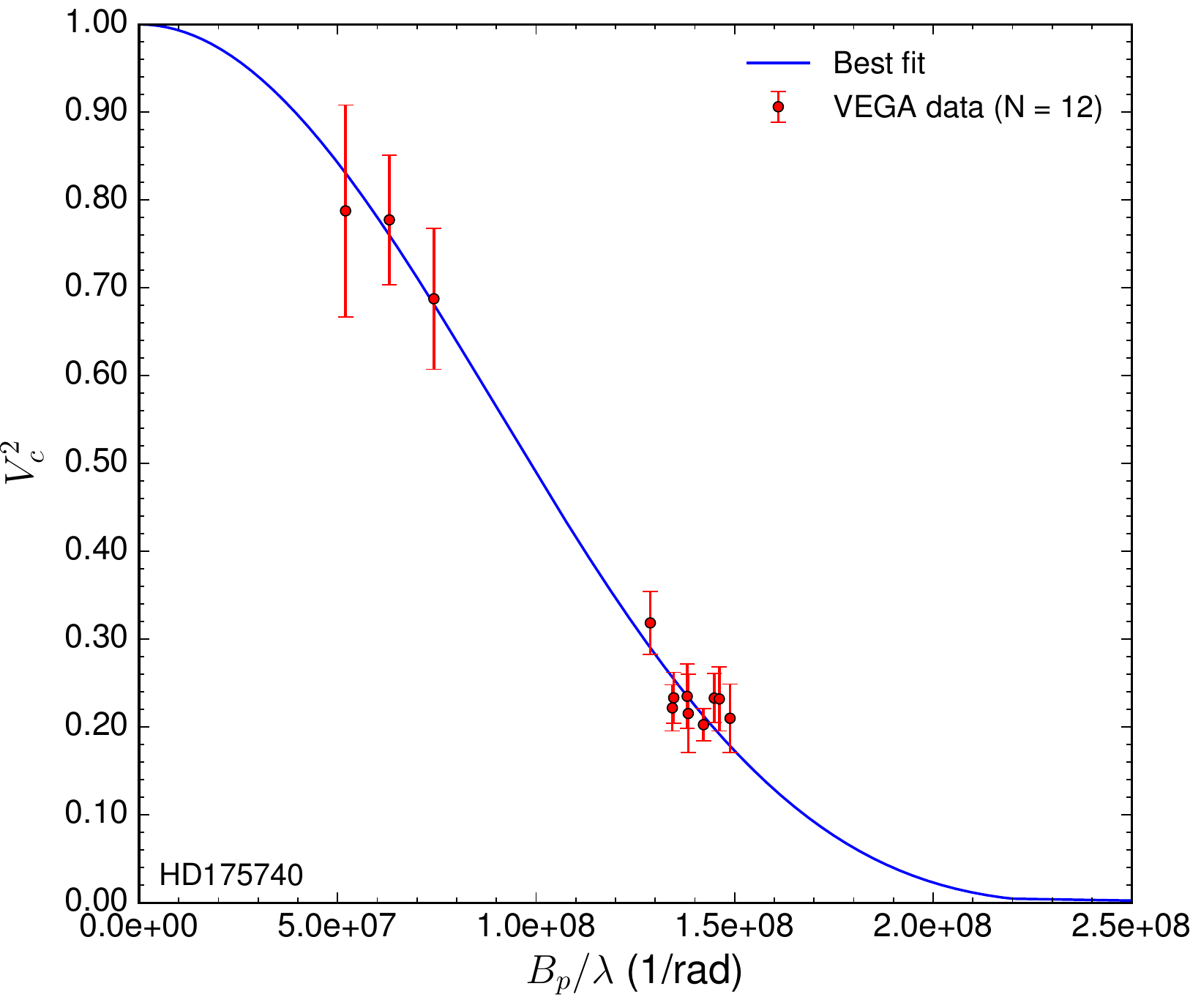}
    \includegraphics[width=0.5\hsize]{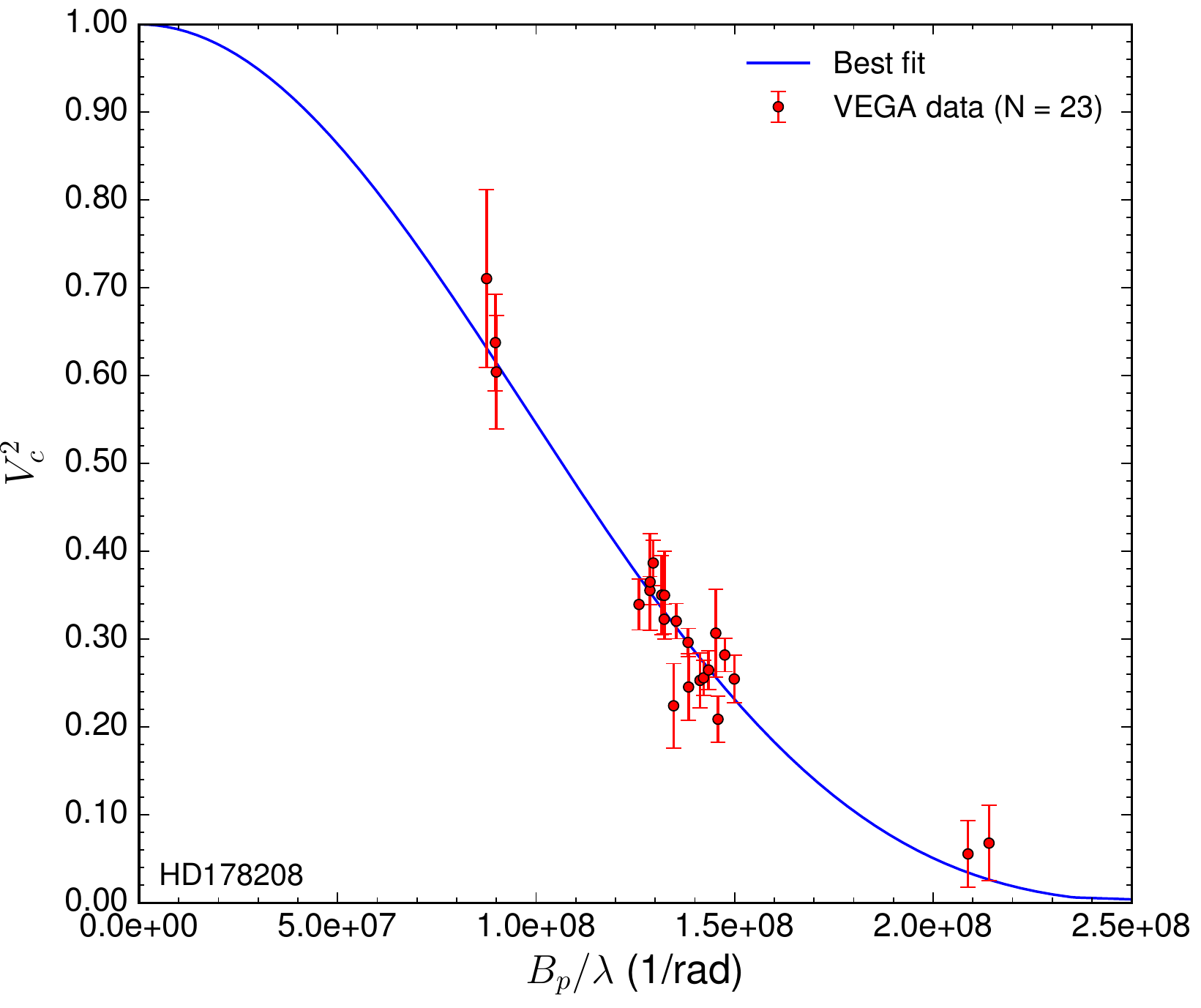}\includegraphics[width=0.5\hsize]{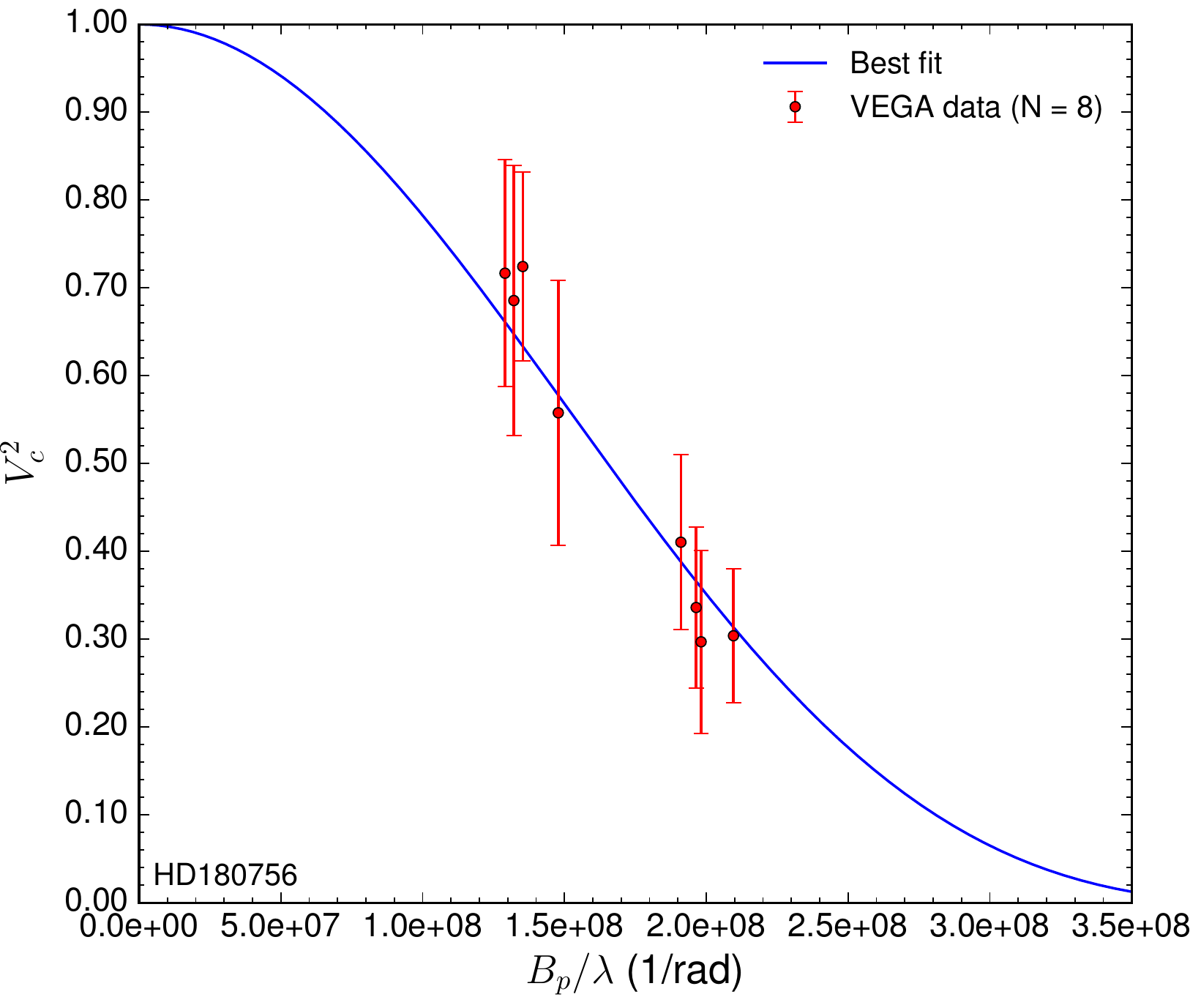} 
    \includegraphics[width=0.5\hsize]{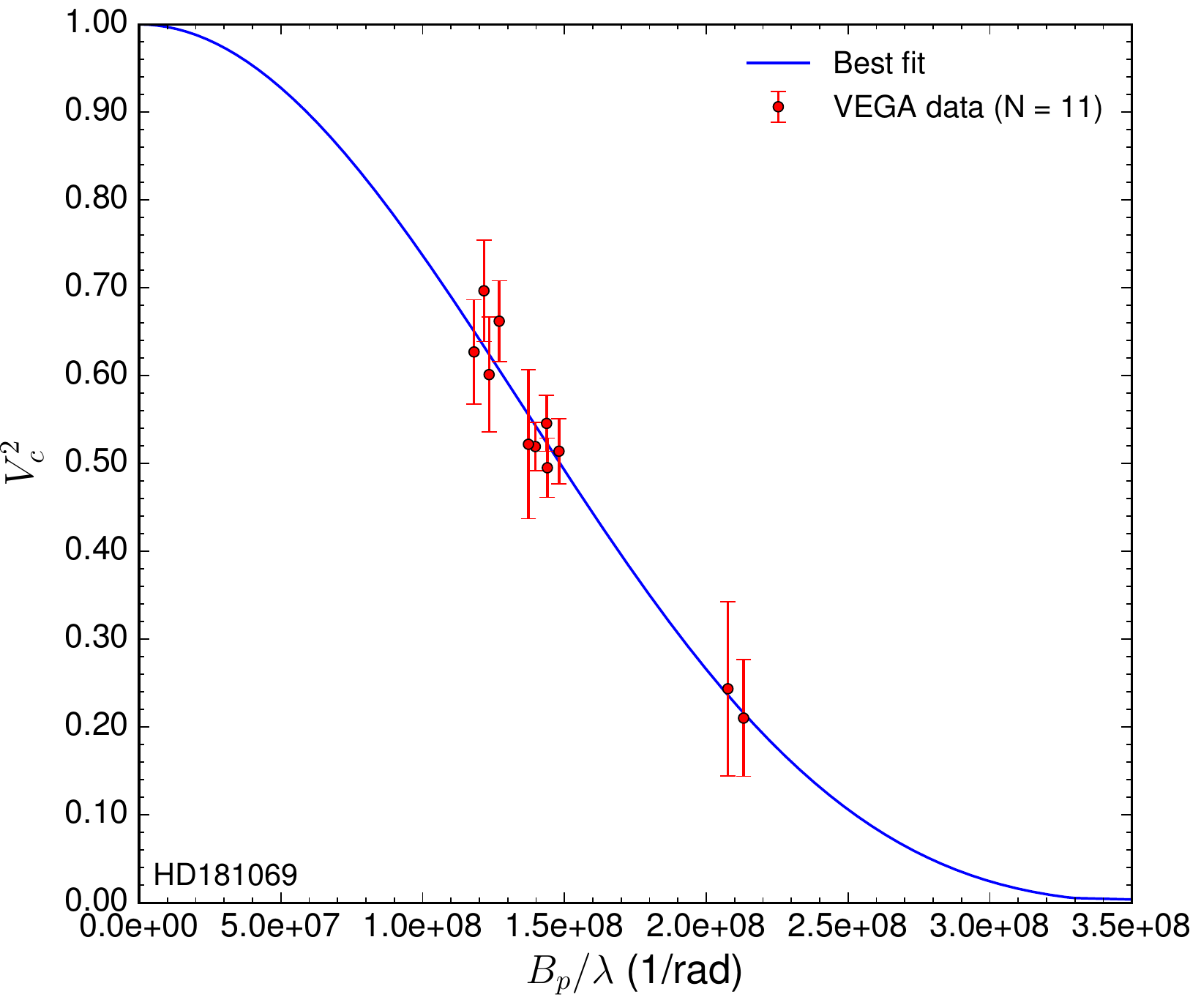}\includegraphics[width=0.5\hsize]{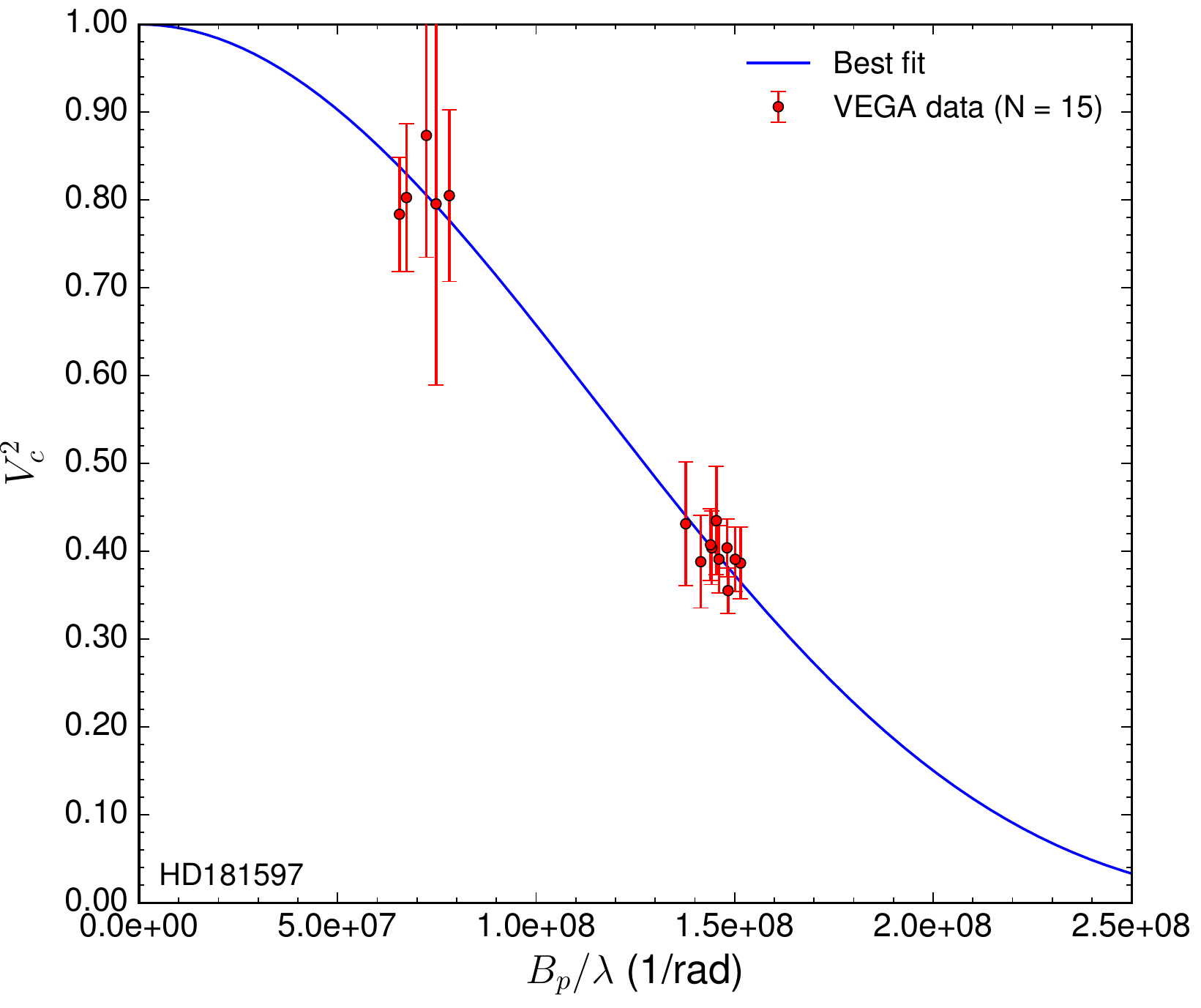}
\end{figure*}
\begin{figure*} \ContinuedFloat  
    \includegraphics[width=0.5\hsize]{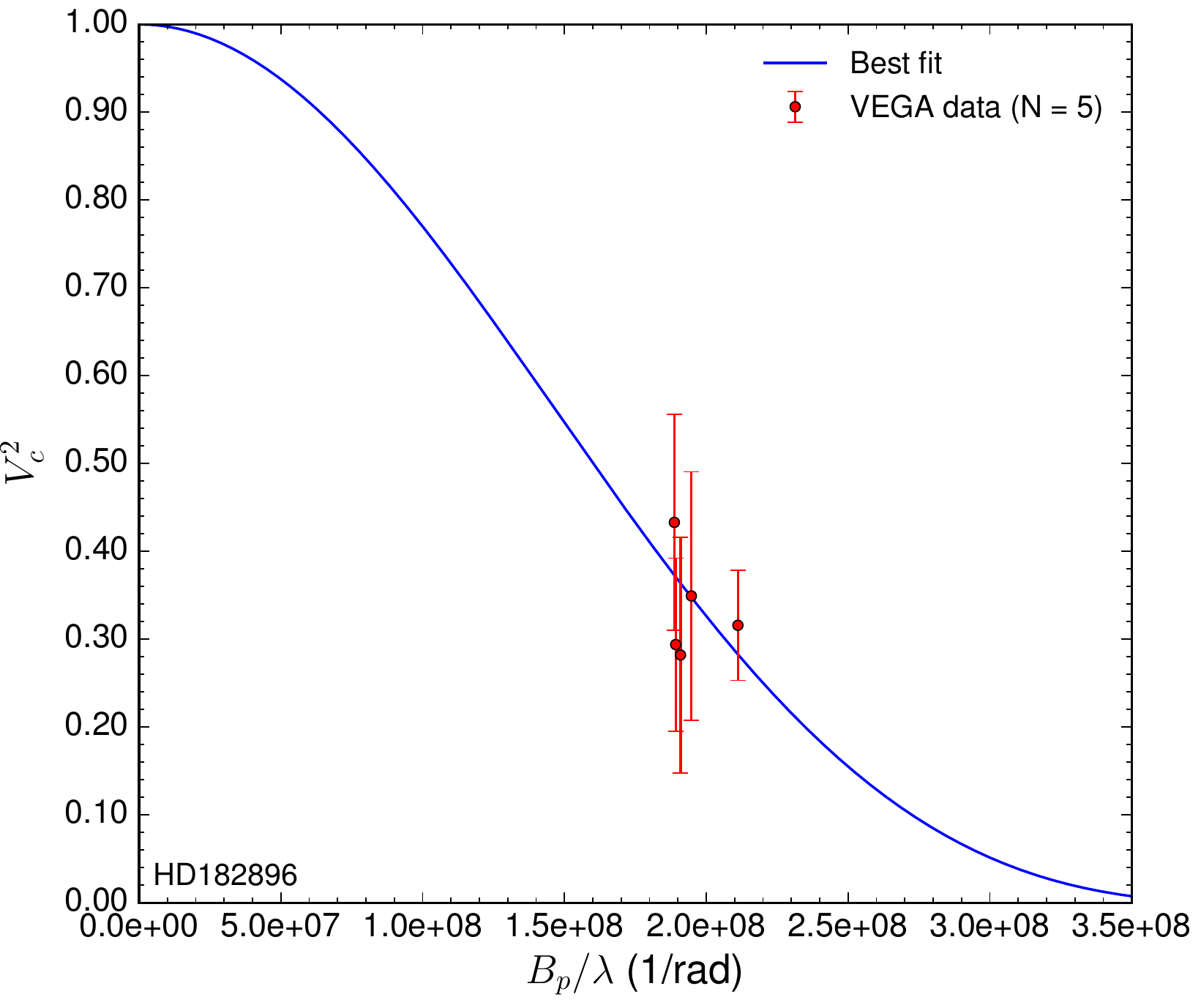}\includegraphics[width=0.5\hsize]{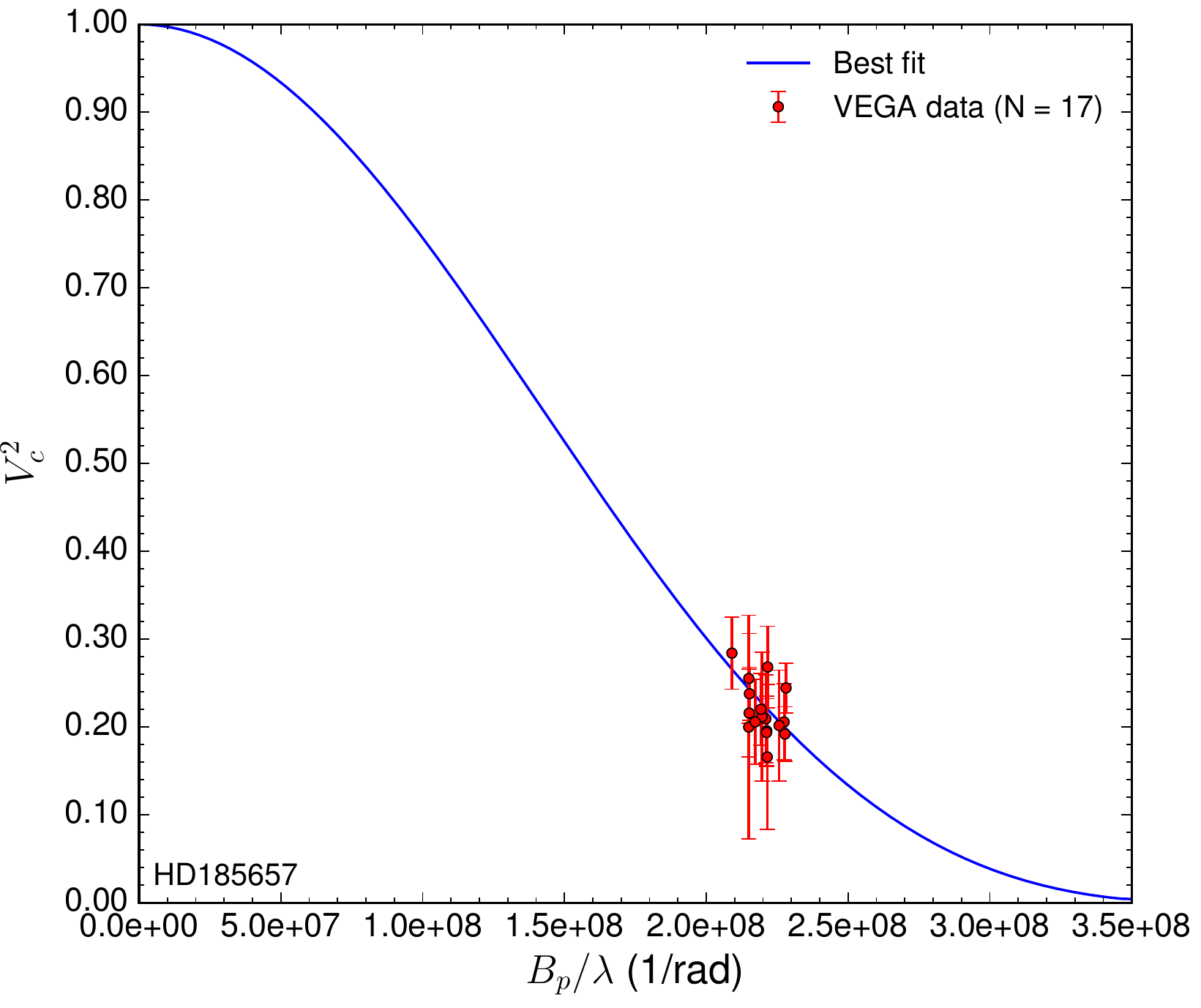}
      \includegraphics[width=0.5\hsize]{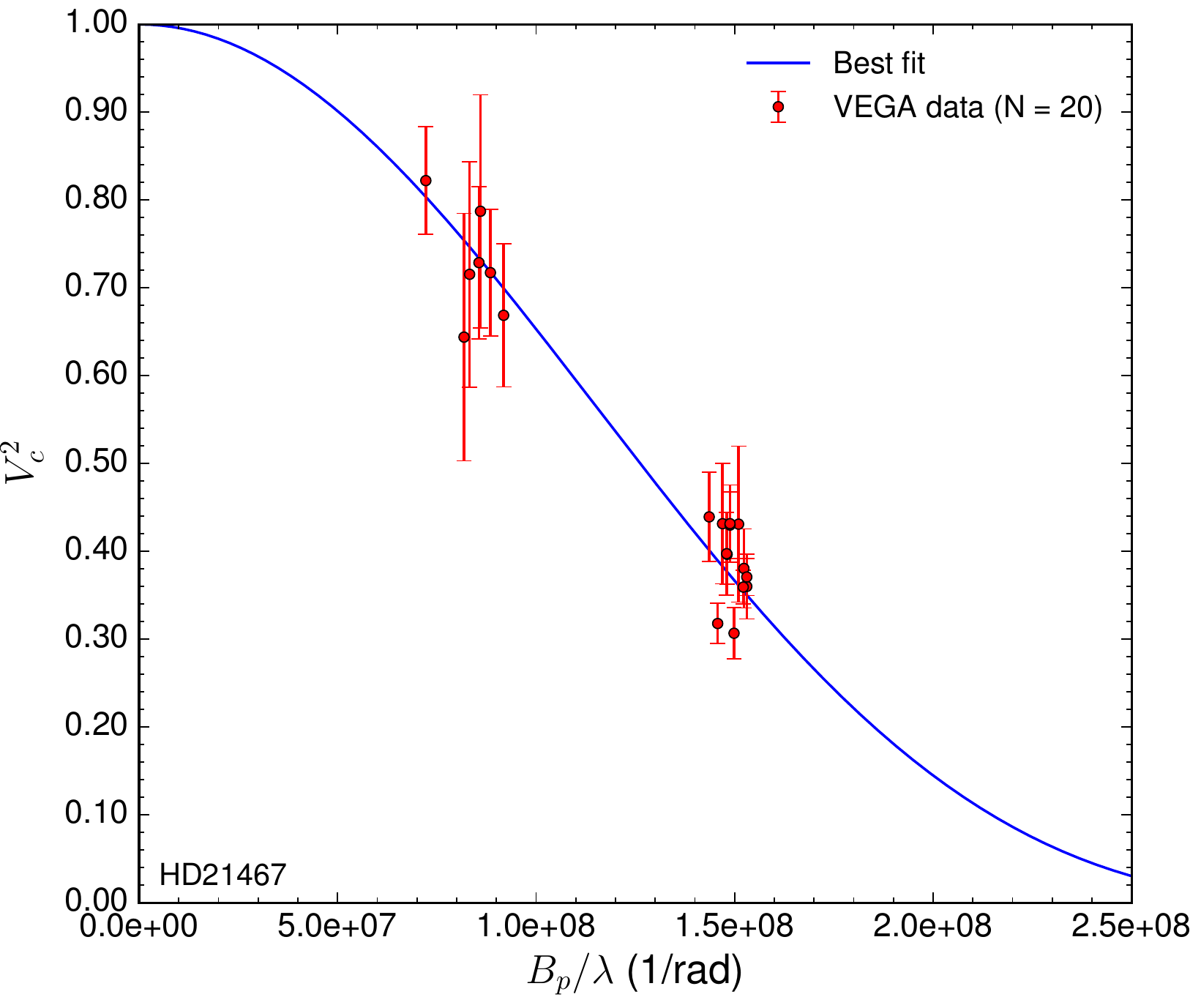}\includegraphics[width=0.5\hsize]{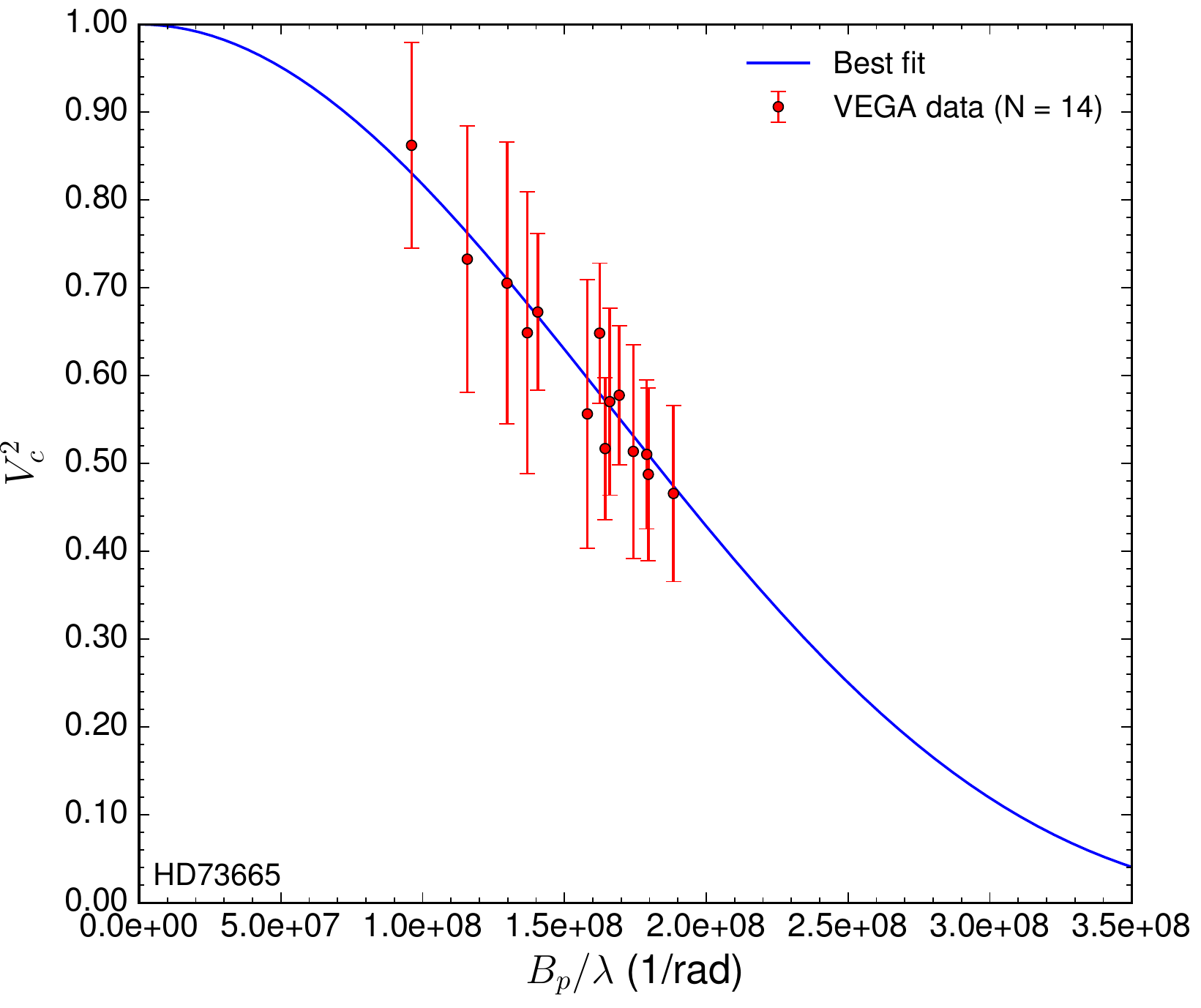}
      \caption{Interferometric squared visibilities of the ten benchmark stars measured by VEGA. The continuous line shows the best fitting model for a limb-darkened disc.}
      \label{vis_fig}

\end{figure*}

\clearpage
\onecolumn

%\longtab[1]{
\begin{longtable}{ccccccccccc}
\caption{\label{obs_log1}VEGA observing log.}\\
\hline
\hline
Star    &       Date    &       Peak    &       AH      &       $\lambda$       &       $\lambda_{\mathrm{min}}$        &       $\lambda_{\mathrm{max}}$        &       $B_p$   &       Arg     &       SNR     &       $V^2_{\mathrm{cal}_{\pm \mathrm{stat} \pm \mathrm{syst}}}$      \\
        &       [yyyy.mm.dd]    &               &       [h]     &       [nm]    &       [nm]    &       [nm]    &       [m]     &       [deg]   &               &               \\
\hline
\endfirsthead
\caption{Continued.}\\
\endhead
\endfoot
HD167042        &       2012.09.21      &       1       &       4.42    &       710     &       700     &       720     &       64.78   &       -179.01 &       24.90   &       0.823$_{\pm 0.033 \pm 0.001}$       \\
        &       2012.09.22      &       1       &       2.32    &       710     &       700     &       720     &       65.55   &       -150.83 &       19.60   &       0.662$_{\pm 0.034 \pm 0.001}$       \\
        &       2012.09.22      &       1       &       2.44    &       730     &       720     &       740     &       65.48   &       -152.38 &       2.75    &       0.777$_{\pm 0.283 \pm 0.001}$       \\
        &       2013.07.29      &       1       &       4.77    &       710     &       700     &       720     &       64.81   &       176.13  &       9.39    &       0.716$_{\pm 0.076 \pm 0.001}$       \\
        &       2013.07.29      &       1       &       4.77    &       730     &       720     &       740     &       64.81   &       176.16  &       7.82    &       0.667$_{\pm 0.085 \pm 0.001}$       \\
\hline                                                                                                                                                                  
HD175740        &       2019.02.26      &       1       &       -5.08   &       710     &       700     &       720     &       36.93   &       -73.11  &       7.13    &       0.788$_{\pm 0.121 \pm 0.001}$       \\
        &       2019.02.26      &       1       &       -4.29   &       710     &       700     &       720     &       44.76   &       -83.32  &       10.55   &       0.777$_{\pm 0.074 \pm 0.001}$       \\
        &       2019.02.26      &       1       &       -3.36   &       710     &       700     &       720     &       52.73   &       -92.88  &       8.56    &       0.688$_{\pm 0.080 \pm 0.001}$       \\
        &       2019.05.04      &       1       &       -2.41   &       730     &       720     &       740     &       94.01   &       -52.38  &       8.92    &       0.319$_{\pm 0.036 \pm 0.001}$       \\
        &       2019.05.04      &       1       &       -1.90   &       730     &       720     &       740     &       98.05   &       -59.11  &       8.46    &       0.222$_{\pm 0.026 \pm 0.001}$       \\
        &       2019.05.04      &       1       &       -1.52   &       730     &       720     &       740     &       100.78  &       -63.76  &       6.43    &       0.235$_{\pm 0.037 \pm 0.001}$       \\
        &       2019.06.12      &       1       &       1.47    &       710     &       700     &       720     &       105.69  &       85.08   &       3.66    &       0.210$_{\pm 0.039 \pm 0.001}$       \\
        &       2019.06.12      &       1       &       1.45    &       730     &       720     &       740     &       105.76  &       85.26   &       6.31    &       0.232$_{\pm 0.036 \pm 0.001}$       \\
        &       2019.06.12      &       1       &       1.85    &       710     &       700     &       720     &       103.77  &       81.14   &       4.02    &       0.233$_{\pm 0.028 \pm 0.001}$       \\
        &       2019.06.12      &       1       &       1.85    &       730     &       720     &       740     &       103.78  &       81.15   &       10.98   &       0.203$_{\pm 0.018 \pm 0.001}$       \\
        &       2019.06.12      &       1       &       2.64    &       710     &       700     &       720     &       98.21   &       72.58   &       4.83    &       0.216$_{\pm 0.045 \pm 0.001}$       \\
        &       2019.06.12      &       1       &       2.63    &       730     &       720     &       740     &       98.31   &       72.72   &       8.08    &       0.233$_{\pm 0.029 \pm 0.001}$       \\
\hline                                                                                                                                                                  
HD178208        &       2017.03.13      &       1       &       -3.21   &       710     &       700     &       720     &       93.97   &       -37.00  &       7.00    &       0.350$_{\pm 0.050 \pm 0.001}$       \\
        &       2017.03.13      &       1       &       -3.22   &       730     &       720     &       740     &       93.96   &       -36.97  &       6.66    &       0.365$_{\pm 0.055 \pm 0.001}$       \\
        &       2017.04.12      &       1       &       -3.61   &       710     &       700     &       720     &       91.95   &       -30.63  &       7.74    &       0.387$_{\pm 0.026 \pm 0.001}$       \\
        &       2017.04.12      &       1       &       -3.61   &       730     &       720     &       740     &       91.92   &       -30.54  &       6.79    &       0.340$_{\pm 0.029 \pm 0.001}$       \\
        &       2017.04.12      &       1       &       -3.23   &       710     &       700     &       720     &       93.91   &       -36.83  &       6.46    &       0.323$_{\pm 0.017 \pm 0.001}$       \\
        &       2017.04.12      &       1       &       -3.22   &       730     &       720     &       740     &       93.93   &       -36.88  &       7.11    &       0.356$_{\pm 0.016 \pm 0.001}$       \\
        &       2017.04.12      &       1       &       -2.83   &       710     &       700     &       720     &       96.07   &       -42.87  &       6.41    &       0.321$_{\pm 0.020 \pm 0.001}$       \\
        &       2017.04.12      &       1       &       -2.83   &       730     &       720     &       740     &       96.09   &       -42.92  &       7.01    &       0.350$_{\pm 0.045 \pm 0.001}$       \\
        &       2017.04.12      &       1       &       -2.45   &       710     &       700     &       720     &       98.27   &       -48.60  &       4.91    &       0.246$_{\pm 0.038 \pm 0.001}$       \\
        &       2017.04.12      &       1       &       -2.44   &       730     &       720     &       740     &       98.28   &       -48.64  &       4.49    &       0.224$_{\pm 0.048 \pm 0.001}$       \\
        &       2017.04.12      &       1       &       -1.96   &       710     &       700     &       720     &       100.94  &       -55.38  &       5.12    &       0.256$_{\pm 0.020 \pm 0.001}$       \\
        &       2017.04.12      &       1       &       -1.96   &       730     &       720     &       740     &       100.94  &       -55.38  &       5.93    &       0.296$_{\pm 0.016 \pm 0.001}$       \\
        &       2017.04.12      &       1       &       -1.52   &       710     &       700     &       720     &       103.13  &       -61.16  &       6.21    &       0.307$_{\pm 0.050 \pm 0.001}$       \\
        &       2017.04.12      &       1       &       -1.52   &       730     &       720     &       740     &       103.13  &       -61.17  &       5.06    &       0.253$_{\pm 0.031 \pm 0.001}$       \\
        &       2017.04.12      &       1       &       -1.15   &       710     &       700     &       720     &       104.74  &       -65.83  &       5.64    &       0.282$_{\pm 0.019 \pm 0.002}$       \\
        &       2017.04.12      &       1       &       -1.16   &       730     &       720     &       740     &       104.72  &       -65.78  &       5.25    &       0.265$_{\pm 0.022 \pm 0.001}$       \\
        &       2017.04.12      &       1       &       -0.66   &       710     &       700     &       720     &       106.44  &       -71.87  &       5.10    &       0.255$_{\pm 0.027 \pm 0.001}$       \\
        &       2017.04.12      &       1       &       -0.66   &       730     &       720     &       740     &       106.45  &       -71.90  &       4.15    &       0.209$_{\pm 0.026 \pm 0.001}$       \\
        &       2017.07.28      &       1       &       2.73    &       710     &       700     &       720     &       152.00  &       30.73   &       1.36    &       0.068$_{\pm 0.043 \pm 0.001}$       \\
        &       2017.07.28      &       1       &       2.73    &       728     &       718     &       738     &       152.00  &       30.73   &       1.11    &       0.056$_{\pm 0.038 \pm 0.001}$       \\
        &       2017.09.16      &       1       &       4.72    &       710     &       700     &       720     &       63.73   &       -3.26   &       11.65   &       0.638$_{\pm 0.055 \pm 0.001}$       \\
        &       2017.09.16      &       1       &       4.72    &       728     &       718     &       738     &       63.73   &       -3.25   &       7.02    &       0.711$_{\pm 0.101 \pm 0.001}$       \\
        &       2017.09.16      &       1       &       5.22    &       710     &       700     &       720     &       63.89   &       -10.22  &       9.36    &       0.604$_{\pm 0.065 \pm 0.001}$       \\
\hline                                                                                                                                                                  
HD180756        &       2017.06.21      &       1       &       -2.25   &       710     &       700     &       720     &       139.47  &       88.52   &       3.67    &       0.336$_{\pm 0.092 \pm 0.003}$       \\
        &       2017.06.21      &       1       &       -2.25   &       730     &       720     &       740     &       139.47  &       88.52   &       4.12    &       0.411$_{\pm 0.100 \pm 0.004}$       \\
        &       2017.06.21      &       1       &       -1.77   &       730     &       720     &       740     &       144.68  &       83.16   &       3.29    &       0.297$_{\pm 0.104 \pm 0.007}$       \\
        &       2017.06.21      &       1       &       -1.31   &       710     &       700     &       720     &       148.81  &       78.04   &       3.98    &       0.304$_{\pm 0.076 \pm 0.002}$       \\
        &       2017.05.13      &       1       &       -3.19   &       730     &       720     &       740     &       94.20   &       -37.30  &       5.76    &       0.717$_{\pm 0.129 \pm 0.001}$       \\
        &       2017.05.13      &       1       &       -2.78   &       730     &       720     &       740     &       96.47   &       -43.66  &       4.45    &       0.686$_{\pm 0.154 \pm 0.001}$       \\
        &       2017.05.13      &       1       &       -2.37   &       730     &       720     &       740     &       98.77   &       -49.64  &       6.75    &       0.724$_{\pm 0.107 \pm 0.001}$       \\
        &       2017.05.13      &       1       &       0.37    &       730     &       720     &       740     &       107.91  &       -83.85  &       3.69    &       0.558$_{\pm 0.151 \pm 0.001}$       \\
\hline                                                                                                                                                                  
HD181069        &       2017.04.15      &       1       &       -1.71   &       710     &       700     &       720     &       147.01  &       78.36   &       11.80   &       0.244$_{\pm 0.099 \pm 0.004}$       \\
        &       2017.04.15      &       1       &       -1.29   &       710     &       700     &       720     &       150.93  &       74.72   &       11.56   &       0.210$_{\pm 0.066 \pm 0.003}$       \\
        &       2017.04.15      &       2       &       -1.23   &       710     &       700     &       720     &       102.03  &       -67.84  &       17.15   &       0.546$_{\pm 0.032 \pm 0.007}$       \\
        &       2017.04.15      &       2       &       -1.24   &       730     &       720     &       740     &       102.00  &       -67.79  &       19.07   &       0.519$_{\pm 0.027 \pm 0.006}$       \\
        &       2017.04.15      &       2       &       -0.69   &       710     &       700     &       720     &       105.14  &       -73.62  &       13.87   &       0.514$_{\pm 0.037 \pm 0.007}$       \\
        &       2017.04.15      &       2       &       -0.69   &       730     &       720     &       740     &       105.10  &       -73.55  &       14.70   &       0.495$_{\pm 0.034 \pm 0.006}$       \\
        &       2017.04.15      &       1       &       -3.09   &       710     &       700     &       720     &       86.35   &       -43.55  &       12.07   &       0.697$_{\pm 0.058 \pm 0.007}$       \\
        &       2017.04.15      &       1       &       -3.10   &       730     &       720     &       740     &       86.23   &       -43.34  &       13.50   &       0.627$_{\pm 0.059 \pm 0.006}$       \\
        &       2017.04.15      &       1       &       -2.67   &       710     &       700     &       720     &       90.17   &       -49.83  &       14.39   &       0.662$_{\pm 0.046 \pm 0.006}$       \\
        &       2017.04.15      &       1       &       -2.68   &       730     &       720     &       740     &       90.11   &       -49.74  &       9.18    &       0.601$_{\pm 0.066 \pm 0.005}$       \\
        &       2017.05.13      &       1       &       -1.49   &       730     &       720     &       740     &       100.23  &       -64.94  &       6.17    &       0.522$_{\pm 0.085 \pm 0.006}$       \\
\hline                                                                                                                                                                  
HD181597        &       2019.03.14      &       1       &       -4.05   &       710     &       700     &       720     &       47.82   &       -78.30  &       8.64    &       0.803$_{\pm 0.084 \pm 0.001}$       \\
        &       2019.03.14      &       1       &       -4.05   &       730     &       720     &       740     &       47.88   &       -78.41  &       10.36   &       0.784$_{\pm 0.065 \pm 0.001}$       \\
        &       2019.03.14      &       1       &       -3.31   &       710     &       700     &       720     &       53.12   &       -87.57  &       3.80    &       0.796$_{\pm 0.206 \pm 0.001}$       \\
        &       2019.03.14      &       1       &       -3.35   &       730     &       720     &       740     &       52.83   &       -87.06  &       6.29    &       0.874$_{\pm 0.139 \pm 0.001}$       \\
        &       2019.03.14      &       1       &       -2.66   &       730     &       720     &       740     &       57.05   &       -94.84  &       8.23    &       0.805$_{\pm 0.098 \pm 0.001}$       \\
        &       2019.05.04      &       1       &       -0.98   &       710     &       700     &       720     &       105.33  &       -68.00  &       13.70   &       0.355$_{\pm 0.026 \pm 0.001}$       \\
        &       2019.05.04      &       1       &       -0.99   &       730     &       720     &       740     &       105.31  &       -67.92  &       9.69    &       0.404$_{\pm 0.042 \pm 0.001}$       \\
        &       2019.05.04      &       1       &       -0.59   &       710     &       700     &       720     &       106.6   &       -72.77  &       10.69   &       0.391$_{\pm 0.037 \pm 0.001}$       \\
        &       2019.05.04      &       1       &       -0.58   &       730     &       720     &       740     &       106.62  &       -72.85  &       10.16   &       0.391$_{\pm 0.038 \pm 0.001}$       \\
        &       2019.05.04      &       1       &       -0.13   &       710     &       700     &       720     &       107.56  &       -78.12  &       9.48    &       0.387$_{\pm 0.041 \pm 0.001}$       \\
        &       2019.06.15      &       1       &       1.69    &       710     &       700     &       720     &       105.14  &       80.90   &       12.33   &       0.404$_{\pm 0.033 \pm 0.001}$       \\
        &       2019.06.15      &       1       &       1.71    &       730     &       720     &       740     &       105.07  &       80.69   &       9.98    &       0.408$_{\pm 0.041 \pm 0.001}$       \\
        &       2019.06.15      &       1       &       2.11    &       710     &       700     &       720     &       103.23  &       75.97   &       7.06    &       0.435$_{\pm 0.062 \pm 0.001}$       \\
        &       2019.06.15      &       1       &       2.10    &       730     &       720     &       740     &       103.28  &       76.08   &       7.37    &       0.388$_{\pm 0.053 \pm 0.001}$       \\
        &       2019.06.15      &       1       &       2.58    &       730     &       720     &       740     &       100.51  &       70.10   &       6.14    &       0.431$_{\pm 0.070 \pm 0.001}$       \\
\hline                                                                                                                                                                  
HD182896        &       2016.07.29      &       1       &       4.08    &       705     &       695     &       715     &       134.60  &       13.01   &       2.10    &       0.282$_{\pm 0.134 \pm 0.004}$       \\
        &       2016.07.29      &       1       &       4.51    &       685     &       675     &       695     &       133.41  &       6.07    &       2.41    &       0.349$_{\pm 0.141 \pm 0.003}$       \\
        &       2016.07.29      &       1       &       4.51    &       705     &       695     &       715     &       133.41  &       6.07    &       2.99    &       0.294$_{\pm 0.098 \pm 0.005}$       \\
        &       2016.07.29      &       1       &       4.90    &       705     &       695     &       715     &       133.08  &       -0.38   &       3.52    &       0.433$_{\pm 0.123 \pm 0.010}$       \\
        &       2016.08.21      &       1       &       2.44    &       685     &       675     &       695     &       144.64  &       36.85   &       5.04    &       0.316$_{\pm 0.063 \pm 0.006}$       \\
\hline                                                                                                                                                                  
HD185657        &       2015.08.30      &       2       &       3.09    &       685     &       675     &       695     &       150.47  &       25.89   &       2.89    &       0.212$_{\pm 0.073 \pm 0.002}$       \\
        &       2015.08.30      &       2       &       4.69    &       685     &       675     &       695     &       147.27  &       2.76    &       4.75    &       0.255$_{\pm 0.051 \pm 0.003}$       \\
        &       2015.08.30      &       2       &       5.23    &       685     &       675     &       695     &       147.37  &       -5.32   &       4.26    &       0.216$_{\pm 0.050 \pm 0.002}$       \\
        &       2015.08.30      &       2       &       5.24    &       705     &       695     &       715     &       147.37  &       -5.36   &       6.92    &       0.284$_{\pm 0.041 \pm 0.002}$       \\
        &       2016.07.29      &       1       &       1.77    &       705     &       695     &       715     &       154.61  &       -136.84 &       5.36    &       0.220$_{\pm 0.041 \pm 0.001}$       \\
        &       2016.07.29      &       1       &       1.77    &       685     &       675     &       695     &       154.61  &       -136.84 &       3.20    &       0.202$_{\pm 0.063 \pm 0.004}$       \\
        &       2016.07.29      &       1       &       2.26    &       705     &       695     &       715     &       153.17  &       -143.03 &       4.26    &       0.206$_{\pm 0.048 \pm 0.005}$       \\
        &       2016.07.29      &       1       &       2.75    &       685     &       675     &       695     &       151.56  &       -149.53 &       5.03    &       0.194$_{\pm 0.039 \pm 0.004}$       \\
        &       2016.07.29      &       1       &       2.75    &       705     &       695     &       715     &       151.56  &       -149.53 &       1.57    &       0.200$_{\pm 0.127 \pm 0.003}$       \\
        &       2016.08.21      &       1       &       1.22    &       685     &       675     &       695     &       155.80  &       -130.14 &       4.76    &       0.206$_{\pm 0.043 \pm 0.001}$       \\
        &       2016.08.21      &       1       &       1.22    &       705     &       695     &       715     &       155.80  &       -130.14 &       4.18    &       0.209$_{\pm 0.050 \pm 0.001}$       \\
        &       2016.08.21      &       1       &       2.70    &       685     &       675     &       695     &       151.74  &       -148.81 &       2.01    &       0.166$_{\pm 0.082 \pm 0.004}$       \\
        &       2016.08.21      &       1       &       2.70    &       705     &       695     &       715     &       151.74  &       -148.81 &       7.91    &       0.238$_{\pm 0.030 \pm 0.005}$       \\
        &       2016.10.01      &       1       &       0.65    &       685     &       675     &       695     &       156.27  &       -123.49 &       8.62    &       0.245$_{\pm 0.028 \pm 0.001}$       \\
        &       2016.10.01      &       1       &       0.63    &       705     &       695     &       715     &       156.27  &       -123.26 &       5.78    &       0.268$_{\pm 0.046 \pm 0.001}$       \\
        &       2016.10.01      &       1       &       1.07    &       685     &       675     &       695     &       156.01  &       -128.29 &       6.19    &       0.192$_{\pm 0.031 \pm 0.001}$       \\
        &       2016.10.01      &       1       &       1.05    &       705     &       695     &       715     &       156.04  &       -128.13 &       4.97    &       0.196$_{\pm 0.039 \pm 0.001}$       \\
\hline                                                                                                                                                                  
HD21467 &       2017.09.17      &       1       &       -0.54   &       705     &       695     &       715     &       104.22  &       -78.86  &       8.61    &       0.430$_{\pm 0.038 \pm 0.001}$       \\
        &       2017.09.17      &       1       &       -0.56   &       725     &       715     &       735     &       104.11  &       -78.75  &       8.67    &       0.439$_{\pm 0.051 \pm 0.001}$       \\
        &       2017.09.17      &       1       &       -0.11   &       725     &       715     &       735     &       106.54  &       -81.58  &       6.30    &       0.432$_{\pm 0.068 \pm 0.001}$       \\
        &       2017.09.17      &       1       &       0.57    &       705     &       695     &       715     &       107.91  &       -85.64  &       7.20    &       0.360$_{\pm 0.037 \pm 0.001}$       \\
        &       2017.09.17      &       1       &       0.57    &       725     &       715     &       735     &       107.92  &       -85.62  &       8.62    &       0.431$_{\pm 0.089 \pm 0.001}$       \\
        &       2017.09.17      &       1       &       0.97    &       705     &       695     &       715     &       107.40  &       -87.91  &       7.61    &       0.381$_{\pm 0.045 \pm 0.001}$       \\
        &       2017.09.17      &       1       &       0.97    &       725     &       715     &       735     &       107.39  &       -87.93  &       7.93    &       0.397$_{\pm 0.047 \pm 0.001}$       \\
        &       2017.10.13      &       1       &       0.15    &       705     &       695     &       715     &       64.73   &       -123.49 &       8.22    &       0.669$_{\pm 0.081 \pm 0.001}$       \\
        &       2017.10.13      &       1       &       0.87    &       705     &       695     &       715     &       62.39   &       -129.25 &       9.93    &       0.717$_{\pm 0.072 \pm 0.001}$       \\
        &       2017.10.13      &       1       &       0.88    &       725     &       715     &       735     &       62.34   &       -129.35 &       5.93    &       0.787$_{\pm 0.133 \pm 0.001}$       \\
        &       2017.10.13      &       1       &       1.35    &       705     &       695     &       715     &       60.35   &       -133.68 &       8.41    &       0.729$_{\pm 0.087 \pm 0.001}$       \\
        &       2017.10.13      &       1       &       1.34    &       725     &       715     &       735     &       60.36   &       -133.66 &       5.57    &       0.715$_{\pm 0.128 \pm 0.001}$       \\
        &       2017.10.17      &       1       &       -0.30   &       705     &       695     &       715     &       105.65  &       -80.40  &       6.14    &       0.307$_{\pm 0.029 \pm 0.001}$       \\
        &       2017.10.17      &       1       &       -0.30   &       725     &       715     &       735     &       105.66  &       -80.41  &       6.36    &       0.318$_{\pm 0.023 \pm 0.001}$       \\
        &       2017.10.17      &       1       &       0.11    &       705     &       695     &       715     &       107.29  &       -82.91  &       7.19    &       0.359$_{\pm 0.019 \pm 0.001}$       \\
        &       2017.10.17      &       1       &       0.11    &       725     &       715     &       735     &       107.29  &       -82.91  &       7.95    &       0.397$_{\pm 0.034 \pm 0.001}$       \\
        &       2017.10.17      &       1       &       0.52    &       705     &       695     &       715     &       107.91  &       -85.32  &       7.42    &       0.371$_{\pm 0.021 \pm 0.001}$       \\
        &       2017.10.17      &       1       &       0.52    &       725     &       715     &       735     &       107.91  &       -85.32  &       8.63    &       0.432$_{\pm 0.044 \pm 0.001}$       \\
        &       2017.12.09      &       1       &       1.89    &       705     &       695     &       715     &       57.70   &       -139.60 &       4.57    &       0.644$_{\pm 0.141 \pm 0.008}$       \\
        &       2017.12.10      &       1       &       -3.64   &       725     &       715     &       735     &       52.36   &       -105.26 &       13.41   &       0.822$_{\pm 0.061 \pm 0.005}$       \\
\hline                                                                                                                                                                  
HD73665 &       2018.04.26      &       1       &       2.51    &       705     &       695     &       715     &       126.21  &       -137.82 &       6.02    &       0.511$_{\pm 0.085 \pm 0.004}$       \\
        &       2018.04.26      &       1       &       2.96    &       705     &       695     &       715     &       119.37  &       -144.11 &       7.31    &       0.578$_{\pm 0.079 \pm 0.005}$       \\
        &       2018.04.26      &       1       &       2.98    &       725     &       715     &       735     &       119.14  &       35.67   &       6.40    &       0.517$_{\pm 0.081 \pm 0.004}$       \\
        &       2018.10.21      &       1       &       -5.69   &       710     &       700     &       720     &       68.22   &       171.29  &       7.35    &       0.862$_{\pm 0.117 \pm 0.001}$       \\
        &       2018.10.21      &       1       &       -8.63   &       710     &       700     &       720     &       82.18   &       84.62   &       4.84    &       0.733$_{\pm 0.151 \pm 0.002}$       \\
        &       2018.10.21      &       1       &       -1.49   &       730     &       720     &       740     &       94.72   &       106.58  &       4.40    &       0.705$_{\pm 0.160 \pm 0.001}$       \\
        &       2018.10.21      &       1       &       -1.03   &       710     &       700     &       720     &       99.82   &       -76.59  &       7.54    &       0.673$_{\pm 0.089 \pm 0.002}$       \\
        &       2018.10.21      &       1       &       -1.02   &       730     &       720     &       740     &       99.95   &       103.32  &       4.04    &       0.649$_{\pm 0.161 \pm 0.001}$       \\
        &       2018.12.16      &       1       &       2.00    &       710     &       700     &       720     &       133.79  &       48.08   &       4.65    &       0.466$_{\pm 0.100 \pm 0.004}$       \\
        &       2018.12.16      &       1       &       2.42    &       710     &       700     &       720     &       127.50  &       43.24   &       4.95    &       0.488$_{\pm 0.099 \pm 0.004}$       \\
        &       2018.12.16      &       1       &       2.44    &       730     &       720     &       740     &       127.22  &       43.02   &       4.22    &       0.514$_{\pm 0.122 \pm 0.003}$       \\
        &       2018.12.16      &       1       &       2.84    &       730     &       720     &       740     &       121.16  &       37.67   &       8.05    &       0.571$_{\pm 0.106 \pm 0.004}$       \\
        &       2018.12.16      &       1       &       3.25    &       710     &       700     &       720     &       115.30  &       31.43   &       5.85    &       0.648$_{\pm 0.080 \pm 0.006}$       \\
        &       2018.12.16      &       1       &       3.24    &       730     &       720     &       740     &       115.40  &       31.55   &       5.27    &       0.557$_{\pm 0.153 \pm 0.004}$       \\
\hline                                                                                                                                                                  
\end{longtable}
%}

\end{appendix}

\end{document}